\documentclass[11pt, a4paper]{article}
\usepackage{jcappub}

\title{Constructing Realistic Szekeres Models from Initial and Final Data}

\author{Anthony Walters}
\author{\& Charles Hellaby}

\affiliation{Department of Mathematics and Applied Mathematics,\\University of Cape Town,\\Rondebosch, 7701,\\ South Africa}

\emailAdd{tony.walters@uct.ac.za}
\emailAdd{charles.hellaby@uct.ac.za}

\abstract{The Szekeres family of inhomogeneous solutions, which are defined by six arbitrary metric functions, offers a wide range of possibilities for modelling cosmic structure. Here we present a model construction procedure for the quasispherical case using given data at initial and final times. Of the six arbitrary metric functions, the three which are common to both Szekeres and Lema\^{\i}tre-Tolman models are determined by the model construction procedure of Krasinski \& Hellaby. For the remaining three functions, which are unique to Szekeres models, we derive exact analytic expressions in terms of more physically intuitive quantities - density profiles and dipole orientation angles. Using MATLAB, we implement the model construction procedure and simulate the time evolution.}


\arxivnumber{}

\begin{document}
\maketitle

\section{Introduction}
Einstein's theory of General Relativity (GR) is widely regarded within the scientific community to be the best current description of gravitational phenomena, having been confirmed by a wide range of observational tests \cite{pound, Shapiro:1971iv, Kramer:2006nb, Biswas:2008cw, Weisberg:2010zz, will}. While this is not to say that it is necessarily a complete or final theory of gravity, as is evident by the numerous attempts to modify it \cite{weyl, Kaluza:1921tu, Akama:1982jy, fradkin, Woodard:2006nt, Clifton:2011jh}, but rather, it does a good job of explaining, and in some cases predicting, observations rather elegantly. It is thus of great importance to fully understand it's behaviour, and one key method is to investigate exact solutions. Assumptions of isotropy together with the Copernican Principle, and more particularly the cosmological principal, point to the Freedmann-Lemaitre-Robertson-Walker (FLRW) geometry as being a good approximation to the behaviour of the Universe on the largest scales, with linear perturbation theory providing the first order corrections. And, in the mainstream of modern cosmology, almost all observations are analysed and interpreted within this framework. However, there are certainly scales on which exact non-linear GR is needed, possibly large scales.

When attempting to specify a realistic model of the Universe, it is important to be able to relate quantities that characterise the model to observables in the real Universe. The traditional approach when studying dynamics is to specify the initial conditions of the system, i.e. initial data at some initial time, say $t_1$, and then evolve it forward in time to some later instant, say $t_2$. In the case of cosmology one may like to take decoupling as $t_1$, and the present as $t_2$, so as to be able to compare one's model with observations. The trouble is that our observations of the last scattering surface can only weakly constrain some of the initial conditions - others must be evolved backwards, from observations at the current time back to $t_1$, which is often not possible. What ends up happening is one assumes an ansatz for the metric functions and attempts to `tweak' them to fit observations. In addition, the evolution is very sensitive to velocity fluctuations, so it is extremely difficult to get a reasonable evolution from a data set at a single time. The ability to specify a model from a combination of both initial and final data is thus of great utility when trying to construct realistic models from observable data.

Perhaps the most popular inhomogeneous exact solution used in cosmology today is the Lemaitre-Tolman (LT) model. First published in 1933 by Lemaitre \cite{Lemaitre:1933, Lemaitre:1933gd, lemaitre:1933bb}, and then by Tolman \cite{Tolman:1934za} and later popularised by Bondi \cite{Bondi:1947av}, it has since found numerous applications in the fields of astrophysics and cosmology. It is a spherically symmetric non-static solution to the Einstein Field Equations (EFEs) with a dust source, and can be thought of as an assembly of concentric spherical mass shells, each with their own evolution. 
Inspired by the work of Bonnor \cite{bonnor:1956}, Krasinski \& Hellaby published a series of papers \cite{Krasinski:2001yi, Krasinski:2003yp, Krasinski:2003cx, Hellaby:2005ut} (Hereafter referred to as Paper I, II, III and IV respectively) in which they developed a procedure for constructing LT models from a combination of initial and final data. In Paper I, the authors proved that any two spherically symmetric density profiles specified on any two constant time slices can be joined by a LT evolution, and exact implicit formulae for the arbitrary functions that determine the resulting LT evolution were given. 
Paper II extended the procedure to determining the LT model which evolves from an initial velocity distribution to a final state that is determined either by a density distribution or a velocity distribution. Numerical examples of the evolution of various structures were investigated, 
and it was found that initial velocity fluctuations have a much bigger effect on the subsequent evolution than initial density fluctuations. 
Paper III applied the methods developed in previous papers to the formation of a galaxy with a central black hole. 
For this investigation, the first of its kind, the tools of GR were essential since it is not possible to describe a black hole in Newtonian gravity or as a perturbation around a FLRW background. 
Paper IV went on to generalise the model construction methods given in the previous three papers, to find the LT model which satisfies any two of the following `boundary conditions': a simultaneous big bang, a homogeneous density or velocity distribution in the asymptotic future, a simultaneous big crunch, a simultaneous time of maximal expansion, a chosen density or velocity distribution in the asymptotic future, only growing or only decaying modes. Some of these new specification methods were used in \cite{Bolejko:2006vw} to model the Shapely Concentration and the Great Attractor. The importance of velocity profiles was again highlighted in \cite{Mustapha:2000bf} where it was demonstrated that an initial over-density can evolve into a void, given a suitable initial velocity profile.

Another interesting family of inhomogeneous exact solutions are those found by Szekeres \cite{szekeres:1975}, in 1975. In general, these models have no symmetries (i.e. no killing-vectors \cite{Bonnor:1977en}) and are defined by six arbitrary metric functions - representing a freedom to rescale the `radial' coordinate and five degrees of freedom to model inhomogeneity. They are perhaps the most sophisticated exact solutions with a dust source, and offer exciting prospects for modelling fairly complex cosmic structures. There are two classes of Szekeres models, the LT-type ($\beta,_z \neq 0$ or Class I) and the Kantowski-Sachs (KS) type ($\beta,_z = 0$ or Class II)\footnote{For a comprehensive review of the two classes of Szekeres models, see \S19.6 in \cite{Plebanski:2006sd} or, for a more historical account, see \S2.4 in \cite{AKrsICM}.} Bonnor \cite{Bonnor:1976zz, Bonnor:1976zza} showed an interior region of LT-type quasispherical Szekeres spacetime can be matched to the exterior Schwarzschild solution, even though the interior metric has no symmetry. Since the Schwarzschild solution does not contain any gravitational radiation, this implies that such Szekeres models do not radiate, and consequently proves the existence of configurations of collapsing dust clouds that have no symmetry and do not produce gravitational waves. Goode \& Wainwright \cite{Goode:1982zz, Goode:1982pg} introduced a different representation of the Szekeres solutions in which many properties of both subfamilies can be considered together\footnote{The KS-Type was later shown to be a regular limit of the LT-Type \cite{Hellaby:1996zz}}. Furthermore, this formulation facilitates the separation of `exact perturbations' from background FLRW dynamics. Recently, it has been used by Ishak \& Peel \cite{Ishak:2011hz} to study the evolution of large scale structure. Meures \& Bruni \cite{Meures:2011ke} recently considered the KS-Type Szekeres solutions with $\Lambda \neq 0$, originally obtained by Barrow \& Stein-Schabes \cite{Barrow:1984zz}, to model an arbitrary initial matter distribution along one line of sight. They re-parametrised the solutions into the Goode \& Wainwright representation, and gave exact solutions for the growing and decaying modes of the metric perturbation, assuming a flat $\Lambda$CDM background.

Within the LT-type class there are three further subclasses, the most popular of which is the `quasispherical' model,  which we focus on here. The geometry of the `quasi-pseudospherical' and `quasiplanar' models is still poorly understood \cite{Hellaby:2007hq}, perhaps due to our lack of understanding of non-spherical gravity, and so, have not been explored for cosmological applications. The quasispherical model can be thought of as a sequence of non-concentric mass shells, each with a density dipole distribution and its own evolution - a generalization of the LT model \cite{Hellaby:2002nx}. Soon after his publication of these solutions, Szekeres used the quasispherical model to study non-spherical gravitational collapse \cite{Szekeres:1975dx}, and since then they have found cosmological application in the study of light propagation \cite{Krasinski:2010rc} as well as structure formation \cite{Bolejko:2006bb}. Bolejko \cite{Bolejko:2006my} used the quasispherical model to study the evolution and interaction of a void with an adjoining galaxy cluster. He found that small voids surrounded by large overdensities evolve much slower than large isolated voids do. And similarly, large voids enhance the evolution of adjacent galaxy superclusters, causing them to evolve much faster than isolated ones. Sussman \& Bolejko \cite{Sussman:2011bp} presented an approach to describing the dynamics of Szekeres models in terms of `quasi-local' scalar variables. In this formulation the field equations and basic physical and geometric quantities are formally identical to their corresponding expressions in the LT model, thus potentially allowing the generalisation of rigorous LT results to the non-spherical Szekeres geometry.  They then used this formalism to investigate small dipole perturbations away from spherical symmetry, showing that such a configuration is in fact stable. In \cite{Bolejko:2010wc} Bolejko \& Sussman used the quasispherical model to construct several elongated supercluster-like structures with underdense regions between them, thus providing a reasonable coarse-grained description of present day cosmic structures. After averaging, the authors found that such a density distribution produced a spherical void profile which is roughly consistent with observations, without the need for dark energy. Also, by considering a non-spherical inhomogeneity, the definition of a ``centre'' becomes much more nuanced, and thus constraints placed by fitting observations on our position relative to this location become less restrictive. Mishra {\it et al} \cite{Mishra:2012vi} calculate the redshift drift equation for an axially symmetric quasispherical Szekeres Swiss-cheese model, along the axis of symmetry. They compare these results to the drift in the $\Lambda$CDM model and some LT models, finding that they are a good discriminator between them. They go on to propose a method to fully constrain all degrees of freedom in an axially symmetric quasispherical Szekeres model from observable quantities. Apparent horizons for the quasispherical Szekeres model have recently been calculated by Krasinski \& Bolejko \cite{Krasinski:2012hv}.

The aim of this paper is to develop an algorithm by which one can construct realistic Szekeres models from some given data on the initial and final time surfaces. This will entail deriving expressions for the six arbitrary metric functions, which will completely define the Szekeres model between the initial and final time, from some physically more intuitive quantities. Here we choose to work with initial and final density data, although it is foreseeable that this approach can be extended to include velocity data. The structure of this paper is as follows: \S \ref{LT} presents the LT model and a summary of the construction procedure; \S \ref{sz} presents the Szekeres model; \S \ref{work} presents the derivation of expressions for the arbitrary functions, and outlines the model construction procedure; \S \ref{numerics} presents the results of the numerical simulations using the procedure outlined in the previous chapter. Conclusions are then presented in \S \ref{conclusions}, along with some comments on future work.
\section{Lema\^{\i}tre-Tolman models} 
\label{LT}
A brief review of the LT model is necessary because (a) many of its functions carry over to the Szekeres models unchanged, and (b) we are generalising a model construction method given for the LT model.

\subsection{The LT metric}
\label{LT_metric}
This metric, in geometric units ($G=c=1$), is
\begin{equation}
 ds^2 = -dt^2 + \frac{R'^2}{1+f}dr^2 + R^2d\Omega^2
 \label{ltmetric}
\end{equation}
where $d\Omega^2 = d\theta^2 + \sin^2\theta d\phi^2$ is the metric of a unit 2-sphere, and $'\equiv \frac{\partial }{\partial r}$. The function $R=R(t,r)$ is known as the areal radius as it is related to the area of constant-$(t,r)$ 2-surfaces, while $f = f(r)$ is an arbitrary function that determines the type of evolution and the local geometry. The dust source is described by an energy-momentum tensor for a pressure-free perfect fluid,
\begin{align}
T^{ab}=\rho u^a u^b,
\label{Tab}
\end{align}
that is  comoving with the coordinate system, such that
\begin{align}
u^a = \delta^a_t .
\label{ua}
\end{align}
Applying the EFEs to the metric yields two expressions - an equation of motion,
\begin{align}
 \dot R^2 = f + \frac{2M}{R} + \frac{\Lambda}{3} R^2 \label{Rdot},
\end{align}
and an expression for the energy density,
\begin{equation}
 \kappa \rho = \frac{2M'}{R^2R'} ,
\label{rho_lt}
\end{equation}
where $\dot{} \equiv \frac{\partial}{\partial t}$ and $M=M(r)$ is another arbitrary function that gives the gravitational mass within a comoving shell of `radius' $r$. One can see that (\ref{Rdot}) is simply the Friedmann equation for dust, except that $f$ and $M$ are functions of $r$. It is evident from (\ref{Rdot}) that $f(r)$ also represents twice the local energy density per unit mass of the dust particles, so it is often written $f(r)=2E(r)$.\\

\noindent The evolution of $R$ depends on the value of $f$. When $\Lambda=0$, the solutions of (\ref{Rdot}), in terms of a parameter $\eta$, are\footnote{Near the origin, where $f\rightarrow0$ and $M\rightarrow0$, the evolution type is determined by the sign of $Rf/M$ or $f/M^{2/3}$}\\
{\bf Hyperbolic} ($f>0$)
\begin{align}
 R &= \frac{M}{f}(\cosh\eta -1) \notag\\
(\sinh\eta - \eta) &= \frac{f^{3/2}(t-t_b)}{M} \label{hyp1}
\end{align}
{\bf Parabolic} ($f=0$)
\begin{align}
 R &= M\frac{\eta^2}{2}\notag\\
\frac{\eta^3}{6} &= \frac{t - t_b}{M}\label{par1}
\end{align}
{\bf Elliptic} ($f<0$)
\begin{align}
 R &= \frac{M}{(-f)}(1-\cos\eta)\notag\\
(\eta-\sin\eta) &= \frac{(-f)^{3/2}(t-t_b)}{M} \label{elip1}
\end{align}
where $t_b=t_b(r)$ is the last arbitrary function - the `bang time', which gives the local time of the initial singularity $t=t_b$ (i.e. when $R=0$ on each worldline). So particle worldlines can emerge from the bang at different times, typically outer spheres before inner ones.  Since $f=f(r)$, it is entirely possible to have adjacent regions of hyperbolic and elliptic evolution. These regions will be connected by a parabolic shell (or extended region) at the boundary, since $f$ is required to be continuous. A nice example of adjacent elliptic and hyperbolic regions is a re-collapsing dust cloud surrounded by an ever-expanding universe \cite{Hellaby:2009vz}. The time reversed parabolic and hyperbolic cases, obtained by writing $(t_b-t)$ instead of $(t-t_b)$, are also valid solutions. \\


\noindent A useful expression for $R'$ is \cite{Hellaby:1985zz}
\begin{align}
R' =  \left( \frac{M'}{M} - \frac{f'}{f} \right)R - \left[ t_b' + \left( \frac{M'}{M} - \frac{3f'}{2f} (t-t_b) \right)  \right] \dot{R} .
\label{Rprime1}
\end{align}
This is valid for all three expressions (\ref{hyp1}), (\ref{par1})\footnote{In (\ref{Rprime1}), one need not set $f'/f=0$ for the parabolic case, as claimed in \cite{bonnor:1985ct}. On the boundary between hyperbolic and elliptical regions, one has $f=0$ and $f' \neq0$. An $f'$ term remains in (\ref{Rprime1}) if the parabolic limit is taken correctly.}, and (\ref{elip1}).

\subsection{Model construction: initial and final density profiles}
\label{LTmodelconstr}
The most obvious way to construct a LT model is to specify the arbitrary metric functions, $M(r)$, $f(r)$ and $t_b(r)$. Alternatively, one makes a coordinate choice and specifies the initial conditions - for example; 
the density $\rho(t_1,M)$ and velocity $\dot{R}(t_1, M)$ distributions at some initial time $t=t_1$.  Since it is not always obvious what density distribution and model evolution will result from a particular choice of initial conditions, there are situations were it is preferable to determine the metric functions from a combination of initial \textit{and} final data.  
In \cite{Krasinski:2001yi} the authors demonstrated that any two spherically symmetric density profiles specified on any two constant time slices can be joined by a LT evolution, and gave exact implicit formulas for the arbitrary functions that define the resulting LT model. Here we present a slight modification to this procedure, which will be useful for the proposed Szekeres model construction procedure presented in \S\ref{work}.



\subsubsection{Coordinate choice}
\label{LTcoordchoice}
The original formulation of \cite{Krasinski:2001yi} made the coordinate choice $\tilde{r}=M(r)$, however, here we find it convenient to make the choice
\begin{equation}
\tilde{r}=R(t_2, r), \label{r=R}
\end{equation}
and thus
\begin{align}
R'|_{t=t_2}=1.
\end{align}
There are a couple of reasons why this choice is preferable. Firstly, the origin limit calculations for $E'/E$, shown in \S\ref{originbehaviour}, are simpler when $R'=1$. Secondly, the choice is less restrictive on the allowable density profiles which produce a finite origin value of $E'/E$\footnote{The reason for this will become clear in \S \ref{solvingEre}, where we derive an expression for $E'/E$ in terms of the density profiles}. And thirdly, $r=M$ does not allow vacuum regions, $M=$ constant, where the $r$ coordinate would be degenerate. Choice (\ref{r=R}) allows one to write the terms with `radial' derivatives instead with respect to $R_2$, as
\begin{align}
\frac{\partial}{\partial r} = \frac{\partial}{\partial R_2}  \Rightarrow R' = \frac{\partial R}{\partial R_2}, \hspace{0.2cm} M' = \frac{\partial M}{\partial R_2}. \label{RMprime}
\end{align}

\subsubsection{Locating the worldlines of constant \boldmath $M$}
\label{init&final}
Our procedure for determining the LT metric functions differs from \cite{Krasinski:2001yi} only in how one finds corresponding values of $R$ and $M$ from the initial and final density profiles (essentially, how one determines $M(r)$ with the choice (\ref{r=R})). 
Suppose the density distributions at the initial instant, $t=t_1$, and the final instant, $t=t_2$, are given by
\begin{align}
\rho_1(R_1)=\rho(t_1, R_1), \hspace{1cm} \rho_2(R_2)=\rho(t_2, R_2), \label{rho(R)}
\end{align}
where
\begin{align}
R_i(r) := R(t_i, r), \hspace{1cm} i = 1,2.
\end{align}
Since $M$ is constant along particle worldlines, corresponding values of $R$ (and $\rho$) at $t_1$ and $t_2$ can be found by rearranging (\ref{rho_lt}) and integrating over $R$. 
One finds
\begin{align}
M(t_i, R_i) = M_{min} + \frac{\kappa}{2} \int_{R_{i\,min}}^{R_i} \rho_{i} R^{*2} dR^*  \hspace{1cm} i=1,2 ,\label{M(R)}
\end{align}
though $M_{min}$ and $R_{i\,min}$ will typically be zero, unless there is a central black hole. 
For definiteness, it is assumed in the following that the final instant is later than the initial instant, and that the final density is smaller than the initial density, at the same M. That is\footnote{Here $\rho(t_2,M)$ is used for $\rho(t_2, R_2(r(M) ) )$ etc.}
\begin{equation}
\rho(t_2,M) < \rho(t_1,M),\hspace{1cm} t_2>t_1
\label{ass1}
\end{equation}
This implies that matter has expanded along every world line, although the analysis can be easily adapted to the collapse situation. As a result of the assumption (\ref{ass1}) one finds $R_2(M)>R_1(M)$ (see Eqn (3.2) of \cite{Krasinski:2001yi}).  
Thus, evaluating (\ref{M(R)}) with (\ref{rho(R)}) allows one to determine the corresponding values of $R_1$ and $R_2$ (see \S\ref{R1/R2sec} for details on the ranges of integration). Once these corresponding values are determined, the procedure for finding $f$ and $t_b$ follows exactly as in \cite{Krasinski:2001yi}. Since we do not wish to duplicate already published work, we refer the reader to the original article for details of the derivation. In summary, there are three main cases and two borderlines which can evolve between $t_1$ and $t_2$. 
The remainder of the procedure, as adapted to the above coordinate choice, is summarised in appendix \ref{appendix}.  This procedure is central to the Szekeres version that follows.
\section{Szekeres models} 
\label{sz}

\subsection{The Szekeres metric}
The LT-type Szekeres metric \cite{szekeres:1975, Szekeres:1975dx} is
\begin{equation}
 ds^2 = -dt^2 + \frac{(R'-R\frac{E'}{E})^2}{\epsilon + f}dr^2 + \frac{R^2}{E^2}(dp^2 +dq^2) \label{metric}
\end{equation}
where $\epsilon=\pm1, 0$. For $\epsilon+1$, the functions $f(r)$ and $R(t,r)$ have the same significance as in the LT model. The function $E(r,p,q)$ may be written
\begin{equation}
 E(r,p,q) = \frac{S}{2} \left[ \left(\frac{p-P}{S}\right)^2 + \left(\frac{q-Q}{S}\right)^2 + \epsilon \right] \label{Erpq}
\end{equation}
where $S=S(r)$, $P=P(r)$ and $Q=Q(r)$ are arbitrary functions, and $S$, $P$ and $Q$ have natural interpretations in the Riemann projection (see \S \ref{sz_reimann_proj}). The dust source is described by the energy-momentum tensor for a pressure-free perfect fluid that is comoving with the coordinates, such that
\begin{align}
T^{ab}=\rho u^a u^b, \hspace{0.2cm} u^a = \delta^a_t.
\label{Tab2}
\end{align}

\noindent Applying the EFEs to the metric yields two expressions. An equation of motion identical to (\ref{Rdot}), which has solutions (\ref{hyp1}) (\ref{par1}) (\ref{elip1}), and an expression for the energy density
\begin{equation}
 \kappa \rho = \frac{2\left[M'-3M\left(\frac{E'}{E}\right)\right]}{R^2\left[R'-R\left(\frac{E'}{E}\right)\right]} .
 \label{density}
\end{equation}
With $\epsilon=+1$, the function $M(r)$ has the same interpretation as in the LT model - it gives the gravitational mass within a comoving shell of `radius' $r$. Any equations from \S\ref{LT} involving only $R$, $M$, $f$, $a$ and $t$, and their derivatives, continue to hold.

\subsection{Quasi-spherical case}
From the family of Szekeres models it is the $\epsilon=+1$ quasi-spherical case that has received the most attention in the field of cosmology. They have found applications in the study of the early Universe \cite{Gron:1985hw, Moffat:2006ct}, structure formation \cite{Bolejko:2006bb, Bolejko:2006my}, the dimming of the supernovae \cite{Bolejko:2010eb, Ishak:2007rp}, light propagation \cite{Krasinski:2010rc}, CMB observations \cite{Bolejko:2008xh} and volume averaging \cite{Bolejko:2008zv}. The $\epsilon \leq 0$ cases have been far less investigated \cite{Hellaby:2007hq}, the geometry is still not well understood, and they are yet to find cosmological application.  While these models may prove of some use in future, in this paper we focus attention on the quasi-spherical case, in which we expect the generalisation of the LT model construction procedure to be easiest.

\subsubsection{Riemann projection}
\label{sz_reimann_proj}
In (\ref{metric}) the metric component $(dp^2 +dq^2)/E^2$ is a unit 2-sphere, plane or pseudo-2-sphere in Riemann projection. When $\epsilon =+1$, the $p$-$q$ 2-surfaces are related to $\theta$-$\phi$ 2-surfaces by one of the following transformations. Either
\begin{align}
\left(\frac{p-P}{S}\right) = \cot \left( \frac{\theta}{2} \right) \cos(\phi) \hspace{1cm}
\left(\frac{q-Q}{S}\right) = \cot \left( \frac{\theta}{2} \right) \sin(\phi) \label{proj+1a}
\end{align}
or
\begin{align}
\left(\frac{p-P}{S}\right) = \tan \left( \frac{\theta}{2} \right) \cos(\phi) \hspace{1cm}
\left(\frac{q-Q}{S}\right) = \tan \left( \frac{\theta}{2} \right) \sin(\phi), \label{proj+1b}
\end{align}
with similar expressions \cite{Hellaby:2007hq} for $\epsilon\leq 0$. The transformed 2-metric is then
\begin{align}
ds^2 =  R^2(d\theta^2 +\sin^2(\theta)d\phi^2 ). \label{2-sphere}
\end{align}
Each of the spherical transformations, (\ref{proj+1a}) and (\ref{proj+1b}), cover the entire $p$-$q$ plane with $0 \leq \theta \leq \pi$ and $0 \leq \phi \leq 2 \pi$, so either can be used. See \cite{Hellaby:2007hq} for illustrations of the various projections.

\subsubsection{The function \boldmath{$E$} and spatial foliations}
\label{Eandsf}
 Since the transformations (\ref{proj+1a}) - (\ref{proj+1b}) are ill-defined if $S=0$, it is normal to assume $S>0$. Then it is clear from (\ref{Erpq}) that when $\epsilon=+1$ we have $E>0$. Differentiating (\ref{Erpq}) with respect to $r$ and setting $E'=0$ gives
\begin{equation}
\left[ p -\left(P - P'\frac{S}{S'}\right) \right]^2 + \left[ q -\left(Q - Q'\frac{S}{S'}\right) \right]^2 = S^2 \left( \frac{P'^2 + Q'^2}{S'^2} + \epsilon \right). \label{E'pq=0locus}
\end{equation}
This is the equation of a circle in the $p$-$q$ plane centred at the point $(p,	q)=(P-P'S/S', Q-Q'S/S')$ with radius $S\sqrt{(P'^2 + Q'^2)/S'^2 + \epsilon}$. When $S'>0$, one finds $E'<0$ inside the circle and $E'>0$ outside \cite{Hellaby:2002nx}.  The locus  (\ref{E'pq=0locus}) has the effect of creating poles and zeros in the function $E'/E$, which appears in $g_{rr}$ and $\rho$, see (\ref{metric}) and (\ref{density}), and thus plays an important role in the geometry and physics of the model. The constant-$(t,r)$ 2-surfaces are multiplied by the factor $R=R(t,r)$, and hence the $r$-$p$-$q$ 3-spaces are constructed from a sequence of constant-$(t,r)$ 2-spheres which have a radius $R$. The $g_{rr}$ component of the metric (\ref{metric}) is affected by $p$-$q$ variations via the function $E'/E$ (\ref{E'E}), which means that the radial separation between two neighbouring spheres of constant-$(t,r)$ also has $p$-$q$ variation, and thus these 2-spheres are interpreted as being arranged non-symmetrically relative to each other.

\subsubsection{\boldmath{$E'/E$} dipole}
\label{eredipole}
Here we show that the function $E'/E$ has a dipole variation around each constant-($t$, $r$) 2-sphere, with the extrema located at antipodal points, and $E'/E=0$ on the `equator' between the two `poles'. In order to illustrate the variation of $E'/E$ over a constant-($t$, $r$) 2-sphere ($\theta$, $\phi$)-coordinates are convenient. Applying either of the spherical transformations (\ref{proj+1a}) or (\ref{proj+1b}) to (\ref{Erpq}) and its derivative, yields
\begin{equation}
 E = \frac{S}{1-\cos\theta} \label{Etp},
\end{equation}
\begin{equation}
 E' = -\frac{S'\cos\theta+\sin\theta(P'\cos\phi + Q'\sin\phi)}{1-\cos\theta} \label{E'tp}
\end{equation}
and
\begin{equation}
 \frac{E'}{E} = -\frac{S'\cos\theta+\sin\theta(P'\cos\phi + Q'\sin\phi)}{S} \label{E'E}.
\end{equation}
The locus $E'/E=0$ is
\begin{equation}
S' \cos\theta + P' \sin\theta \cos \phi + Q' \sin \theta \sin \phi = 0 \label{E'=0locus}
\end{equation}
and applying the rectangular transformations
\begin{equation}
x= \sin \theta \sin \phi, \hspace{0.5cm} y=\sin\theta\cos\phi \hspace{0.5cm} z=\cos\theta \label{rect}
\end{equation}
allows for a natural interpretation
\begin{equation}
P'x + Q'y + S'z=0 \label{plane0}
\end{equation}
which is the equation of an arbitrary plane passing through $(0,0,0)$. The locus $E'/E=0$ is then the intersection of this plane with the unit 2-sphere, which is a great circle. The unit normal to the plane (\ref{plane0}) is
\begin{equation}
\vec{n} =  \frac{(P',Q',S')}{\sqrt{(P')^2 + (Q')^2 + (S')^2}}. \label{normal}
\end{equation}
By setting (\ref{E'E}) equal to a constant and employing the rectangular transformations mentioned above, one finds the loci $E'/E=$constant
\begin{equation}
P'x + Q'y + S'z=kS \hspace{1cm} k = constant \label{plane1}
\end{equation}
to be the equation of arbitrary planes parallel to (\ref{plane0}). This implies that all the loci $E'/E=$constant are small circles parallel to the $E'=0$ great circle (\ref{E'=0locus}). The location on the 2-sphere of the $E'/E$ extrema (i.e. the poles) are found by finding where the partial derivatives of $E'/E$, with respect to $\theta$ and $\phi$, are equal to zero. Denoting the location of the extrema by $(\theta_e, \phi_e)$ one finds
\begin{align}
 \frac{\partial(E'/E)}{\partial\phi} &= \frac{\sin\theta(P'\sin\phi-Q'\cos\phi)}{S}=0 \notag\\
\Rightarrow \tan\phi_e &= \frac{Q'}{P'} \label{tanphie}\\
\Rightarrow \cos\phi_e &= \epsilon_1 \frac{P'}{\sqrt{P'^2+Q'^2}}, \hspace{1cm} \epsilon_1 = \pm1 \label{cosphie}
\end{align}

\begin{align}
 &\frac{\partial(E'/E)}{\partial\theta} = \frac{S'\sin\theta-P'\cos\theta\cos\phi - Q'\cos\theta\sin\phi}{S}=0 \notag\\
&\Rightarrow \tan\theta_e = \epsilon_1\frac{\sqrt{P'^2+Q'^2}}{S'} \label{tanthetae}\\
&\Rightarrow  \cos\theta_e = \epsilon_2 \frac{S'}{\sqrt{S'^2+P'^2+Q'^2}},  \hspace{1cm} \epsilon_2 = \pm1 \label{costhetae}
\end{align}
Applying the transformation (\ref{rect}) to the expressions found above gives the location of the extrema in rectangular coordinates. One finds
\begin{align}
(x_e, y_e, z_e) = \epsilon_{2} \frac{(P',Q',S')}{\sqrt{(P')^2 + (Q')^2 + (S')^2}} \label{extreme_vec}
\end{align}
This vector points in the same direction as the vector normal to the plane of the great circle (\ref{normal}), so the extrema of $E'/E$ are located at the poles of $E'/E=0$ great circle. The extreme values of $E'/E$ are found by substituting the location of the extrema (\ref{cosphie}) (\ref{costhetae}) into (\ref{E'E}), which gives
\begin{equation}
 \left(\frac{E'}{E}\right)_{extreme} = -\epsilon_2 \frac{\sqrt{S'^2+P'^2+Q'^2}}{S} \label{ErEext}
\end{equation}
The parameters $\epsilon_2$ and $\epsilon_1$ are not independent\footnote{The presentation in \cite{Hellaby:2002nx} misses this point}. The relationship between the two can be found by relating the sign of (\ref{tanthetae}) to that of (\ref{costhetae}), as follows. Noting that $\theta$ is defined on the interval $[0, \pi]$, we have $\sin \theta \geq 0$, and hence
\begin{align}
 {\rm sign}(\cos \theta_e) &= {\rm sign}(\tan \theta_e) \notag\\
\Rightarrow {\rm sign}(\epsilon_2 S') &=  {\rm sign}(\epsilon_1 / S' ) \notag\\
\Rightarrow \epsilon_1 &= \epsilon_2. \label{e1=e2}
\end{align}
This seems reasonable since we expect only two extrema. From (\ref{ErEext}) and (\ref{e1=e2}), and assuming that $S>0$ always holds, we deduce that $\epsilon_1=\epsilon_2=1$ corresponds to a dipole minimum, while the dipole maximum is given by $\epsilon_1=\epsilon_2=-1$. From this, it is clear that
\begin{align}
 \left(\frac{E'}{E}\right)_{max} = -\left(\frac{E'}{E}\right)_{min}, \label{antisym}
\end{align}
and at the dipole maximum, the $E'/E$ value is then 
\begin{align}
 \left(\frac{E'}{E}\right)_{max} &=  \frac{\sqrt{S'^2+P'^2+Q'^2}}{S} ,
 \label{eremax}
\end{align}
while the orientation angles are
\begin{align}
\cos\phi_{max} &= -\frac{P'}{\sqrt{P'^2 + Q'^2}} \label{cosphimax}\\
\cos\theta_{max} &= -\frac{S'}{\sqrt{S'^2 + P'^2 + Q'^2}} \label{costhetamax}
\end{align}
Sketches of a dipole and the variation of its orientation can be found in section 4.4 of \cite{Hellaby:2009vz}.

%
%

\subsubsection{Shell separation}
As pointed out in \S \ref{Eandsf}, the $g_{rr}$ component of the metric (\ref{metric}) is sensitive to $p$-$q$ (and hence $\theta$-$\phi$) variations via the function $E'/E$, and hence the constant-$(t,r)$ 2-surfaces are interpreted as being arranged non-symmetrically. Writing the radial separation as
\begin{equation}
R' - R\frac{E'}{E} =R' + R \frac{S' \cos\theta + P' \sin\theta \cos \phi + Q' \sin \theta \sin \phi}{S}
\end{equation}
we see that $RE'/E$ is a correction term to the spherically symmetric radial separation, $R'$, that an LT model would have. The minimum radial separation between neighbouring constant-$r$ shells obviously occurs where $E'/E$ is maximum. Hence, $\epsilon=+1$ Szekeres 3-spaces are interpreted as being constructed from a sequence of non-concentric 2-spheres, with each shell having the exact density distribution required to generate a spherical field around the new centre. The dipole variation in $E'/E$ around each constant-$(t,r)$ 2-sphere causes a dipole variation in the density (\ref{density}). At the equator, where $E'/E=0$, we see that (\ref{density}) reduces to the form
\begin{equation}
\kappa\rho_{LT} = \frac{2M'}{R^2R'} \label{lt_density}
\end{equation}
which is identical to the expression for density in LT models (\ref{rho_lt}). In the context of Szekeres models, we refer to the equatorial density (\ref{lt_density}) as the LT-density, denoted by the subscript LT.

\subsection{Singularities}
Szekeres models contain the same singularities as in LT - those of the bang, the crunch and shell crossings. The Kretschmann scalar is
\begin{align}
 K= \kappa^2 \left( \frac{4}{3}\rho_{AV}^2 - \frac{8}{3}\rho_{AV}\rho + 3 \rho^2 \right) + \frac{4\Lambda}{3} \left(2\Lambda + \kappa\rho\right) \label{K2}
\end{align}
where
\begin{equation}
\kappa\rho_{AV} \equiv \frac{6M}{R^3} \label{rho_av}
\end{equation}
is like a ``mean density''\footnote{see e.g. \cite{Hellaby:1996zz}, also c.f. the quasi-local average of \cite{Sussman:2011bp}} interior to the comoving shell of constant-$r$. The dynamics of $R$ in the Szekeres model is identical to that of the LT model (\ref{Rdot}), and hence the bang and crunch singularities are the same. Shell crossings occur when inner shells of matter pass outer ones, causing the density to diverge and the radial coordinate, $r$, to become degenerate. In Szekeres models they are more complicated than in the spherically symmetric LT case, as non-concentricities cause constant-$r$ shells to pass through each other gradually, and one shell may intersect many others at any given time. 
So to avoid shell crossing one requires, in addition to the LT conditions, further constraints on the radial derivatives of the arbitrary functions $S'$, $P'$ and $Q'$. Necessary and sufficient conditions to completely avoid shell crossings in Szekeres models are given in \cite{Hellaby:2002nx} for the $\Lambda=0$ case.

\subsection{Regularity conditions}
\label{reg_cond}
For the metric (\ref{metric}) to retain Lorentzian signature $(-+++)$, the $g_{rr}$ component must always remain positive, and thus $\epsilon + f \geq 0$ is required, with the equality only occurring where $R'-RE'/E=0$. 
In the $\epsilon=+1$ case the origin is the locus $r=r_{0}$ where the 2-sphere radius vanishes, i.e. $R(t, r_{0}) = 0 \hspace*{0.2cm} \forall \: t$ so that $\dot{R}(t,r_0)=0$, $\ddot{R}(t,r_0)=0$ etc. This is not a centre of symmetry, but a locus where the shear and electric Weyl tensors vanish (assuming a regular origin). Regular origins require that on any constant $t$ surface away from the bang or crunch, the density (\ref{density}) and curvature (\ref{K2}) remain finite, and the time evolution at $r=r_{0}$ should be a smooth continuation of the immediate neighbourhood. Conditions on the arbitrary functions were investigated in \cite{Hellaby:2002nx} by evaluating the limit $r\rightarrow r_{0}$ of the aforementioned quantities and ensuring they are well behaved. They found the regularity conditions require that near the origin
\begin{align}
&M \sim R^3 \hspace*{1cm} f \sim R^2 \notag\\
&S \sim R^n \hspace*{1cm} P \sim R^n \hspace*{1cm} Q \sim R^n \hspace{1cm} 0 \leq n \leq 1
\end{align}
Of the five conditions above, the first two are the same as in LT models.
Spatial extrema in $R$ may occur in the constant $t$ 3-spaces of models with certain topologies. For example, closed spatial sections can have $R$ increasing away from the origin until some point, say $r=r_m$, where $R$ is maximal and beyond which $R$ is decreasing toward a second origin. The conditions to ensure regularity of these loci were investigated in \cite{Hellaby:2002nx}. Ensuring no shell crossings at the maximum, $R'=0$ requires
\begin{align}
&M'(t, r_m) = f'(t,r_m) = t_B'(t, r_m) = 0 \notag\\
&S'(t,r_m) = P'(t,r_m) = Q'(t,r_m) = 0 \hspace*{1cm} \forall \: t
\end{align}
Furthermore, to avoid any surface layers at $r=r_m$ requires
\begin{equation}
f = - \epsilon
\end{equation}
With these conditions met, the density (\ref{density}) and the $g_{rr}$ component of the metric (\ref{metric}) remain positive and finite, thus ensuring regular extrema.

\section{Towards a model construction procedure} 
\label{work}


In order to construct a {\it realistic} Szekeres model one must specify the arbitrary functions associated with the model from some physical quantities. These functions, of which there are six in total, allow for a rescaling of the $r$ coordinate, $\tilde{r}=\tilde{r}(r)$, plus five physical degrees of freedom to model inhomogeneity. 
Once one has obtained expressions for these six functions the Szekeres model will be completely specified. Special attention must be paid to the origin as certain variables have a value of zero there.


\subsection{Obtaining the arbitrary functions}
Since the expression for the Szekeres equatorial density (\ref{lt_density}) corresponds exactly with the LT density expression (\ref{rho_lt}), and the dynamics of $R$ are identical in both Szekeres and LT models (\ref{Rdot}), the LT model construction procedure of Krasinski \& Hellaby can be used to determine the arbitrary functions which are common in both models. Thus, specifying a Szekeres equatorial density profile (hereafter the ``LT-density'') at some initial and final time, $t_1$ and $t_2$, is sufficient to determine the arbitrary functions $M$, $f$ and $t_b$ according to the procedure described in \S \ref{LTmodelconstr}. Note however that the radial coordinate choice is not as in \cite{Krasinski:2001yi}. The arbitrary functions $S$, $P$ and $Q$ require further knowledge of the intensity and orientation of the dipole. These dipole parameters can be specified on only one of the 3-surfaces, $t_1$ or $t_2$ - we choose to do it at the later time as we expect one would know more about density detail at later times. Defining the density extrema on a particular constant $r$ shell at time $t_2$ to be
\begin{align}
\rho_{min}(r) \equiv \min _{\theta, \phi}  \left[\rho(t_2, r, \theta, \phi)\right]
\end{align}
and
\begin{align}
\rho_{max}(r) \equiv \max _{\theta, \phi}  \left[\rho(t_2, r, \theta, \phi)\right]
\end{align}
we specify the following:
\begin{itemize}
 \item $\rho_{LT,1}(R_1)$ LT-density profile at $t=t_1$
 \item $\rho_{LT,2}(R_2)$ LT-density profile at $t=t_2$
 \item $\rho_{min}(R_2)$ Density minimum profile at $t=t_2$
 \item $\theta_{\rho_{min}}(R_2)$ Density minimum orientation angle, $\theta$, at $t=t_2$
\item $\phi_{\rho_{min}}(R_2)$ Density minimum orientation angle, $\phi$, at $t=t_2$
\end{itemize}
From these quantities we will extract expressions for the arbitrary metric functions, $S$, $P$ and $Q$.


\subsubsection{Density extrema}
\label{miscon}
Previous literature (\cite{Hellaby:2002nx} - Equation 69)  claimed to show the derivative of the density with respect to $E'/E$ to be negative (i.e. $\rho_{,x}<0$ with $x=E'/E$), implying that $E'/E$ is a minimum at a density maximum, and a maximum at a density minimum. Moreover, since $E'/E$ is a dipole (see \S \ref{eredipole}), it implies that a density maximum corresponds to a negative $E'/E$ value, and a density minimum to a positive $E'/E$ (i.e. $E'/E|_{\rho_{max}}=E'/E|_{min}\leq0$ and $E'/E|_{\rho_{min}}=E'/E|_{max}\geq0$). This is in fact not the case. Taking the derivative of the density (\ref{density}) with respect to $x=E'/E$, and using (\ref{lt_density}) and (\ref{rho_av}), we find
\begin{align}
 \rho,_{x} &= R R' \left[ \frac{(\rho_{LT}-\rho_{AV})}{ (R'-Rx)^2} \right] .
\label{rhox}
\end{align}
Clearly the sign of $\rho,_x$ is not always negative, as was previously thought, but rather it depends on the sign of $R'(\rho_{LT}-\rho_{AV})$, since $R/(R'-Rx)^2>0$ always holds. Thus, $E'/E|_{max}$ only occurs at the density minimum if $R'(\rho_{LT}-\rho_{AV})<0$. 

\subsubsection{Solving for \boldmath{$E'/E$}}
\label{solvingEre}
In order to express $E'/E$ in terms of density parameters,  we begin by rearranging (\ref{density}) to give
\begin{align}
 \left(\frac{E'}{E}\right) &= \frac{\kappa\rho R^2 R' - 2M'}{\kappa\rho R^3 -6M} .
\label{ere1}
\end{align}
Then, dividing both numerator and denominator by $\kappa R^3$ and substituting in (\ref{rho_av}) and (\ref{lt_density}), leads to
\begin{align}
\left(\frac{E'}{E}\right) 
&= \frac{R'}{R} \left( \frac{\rho - \rho_{LT}}{\rho-\rho_{AV}} \right).
\label{ere}
\end{align}
The anti-symmetric property of $E'/E$ (\ref{antisym}) allows one to express $E'/E|_{max}$ in terms of $E'/E|_{\rho_{min}}$, regardless of the sign of $R'(\rho_{LT}-\rho_{AV})$. 
Substituting (\ref{ere}), with $\rho=\rho_{min}$ , into (\ref{antisym}) gives
\begin{align}
 \left( \frac{E'}{E} \right)_{max} & =  \left|\frac{R'}{R} \left( \frac{\rho_{min} - \rho_{LT}}{\rho_{min} - \rho_{AV}} \right) \right| .\label{E'Emax}
\end{align}
However, when relating the orientation angles of the density-dipole to the $E'/E$ dipole, it is necessary to track the sign of $R'(\rho_{LT}-\rho_{AV})$. Since the poles of $E'/E$ (and $\rho$) are antipodes on the 2-sphere, the orientation angles are easily related. 
\begin{align}
&R'(\rho_{LT}-\rho_{AV}) > 0: \hspace{1cm} &\theta_{max}&= \pi - \theta_{\rho_{min}}, \hspace{1cm} &\phi_{max} =& \pi + \phi_{\rho_{min}}  \label{cs1}\\
&R'(\rho_{LT}-\rho_{AV}) < 0: \hspace{1cm} &\theta_{max}&= \theta_{\rho_{min}}, \hspace{1cm} &\phi_{max} =& \phi_{\rho_{min}} .  \label{cs2}
\end{align}
We now have expressions for $E'/E|_{max}$, $\theta_{max}$, $\phi_{max}$ as functions of $r$ in terms of the physical quantities $\rho_{LT}$, $\rho_{AV}$, $\rho_{min}$, $\theta_{\rho_{min}}$ and $\phi_{\rho_{min}}$ at time $t_2$. We can thus go on to define the arbitrary functions, $S$, $P$ and $Q$, using these expressions.

\subsubsection{Solving for \boldmath{$S$}, \boldmath{$P$} and \boldmath{$Q$}}
\label{solvingspq}
Equations (\ref{cosphimax}), (\ref{costhetamax}) and (\ref{eremax}) suggest that the radial derivatives of the three arbitrary functions, $S'$, $P'$ and $Q'$, can be solved for in terms of the dipole orientation angles, $\theta_{max}$ and $\phi_{max}$, and  $E'/E_{max}$. The metric functions, S P and Q, can then found by integrating the expressions for $S'$, $P'$ and $Q'$ over a suitable radial coordinate. Solving the system of equations (\ref{eremax}) (\ref{cosphimax}) (\ref{costhetamax}), we find the following. Multiplying (\ref{eremax}) by $S$, squaring both sides and rearranging leads to
\begin{align}
S' = \pm \sqrt{\left( \frac{E'}{E} \right)_{max}^2S^2 - P'^2 - Q'^2}, \label{z}
\end{align}
and substituting this expression (\ref{z}) into (\ref{costhetamax}), and using (\ref{eremax}), eliminates the $S'$ terms, to give
\begin{align}
 \cos \theta_{max} = \frac{\mp\sqrt{\left( \frac{E'}{E} \right)_{max}^2S^2 - P'^2 - Q'^2}}{\left( \frac{E'}{E} \right)_{max}S}. \label{temp2}
\end{align}
Multiplying (\ref{temp2}) by the RHS denominator, squaring and rearranging yields an expression for $Q'^2$ in terms of only $P'$ and $S$,
\begin{align}
Q'^2 &= \left( \frac{E'}{E} \right)_{max}^2S^2\sin \theta_{max}^2-P'^2. \label{y2}
\end{align}
Now $Q'^2$ can be entirely eliminated from (\ref{cosphimax}) by substituting it into (\ref{y2}), 
and rearranging the result then gives an expression for $P'$,
\begin{equation}
P' = -\sin \theta_{max} \cos \phi_{max}  \left( \frac{E'}{E} \right)_{max}S . \label{x}
\end{equation}
By substituting (\ref{x}) into (\ref{cosphimax}) one eliminates $P'$, and solving for $Q'$ one finds
 \begin{align}
Q' = \epsilon_3 \sin \theta_{max} \sin \phi_{max} \left( \frac{E'}{E} \right)_{max}S \hspace*{1cm} \epsilon_3 = \pm 1 \label{y}
\end{align}
From (\ref{eremax}) and (\ref{costhetamax}) it is easily seen that
\begin{equation}
 S' = -\cos \theta_{max}\left( \frac{E'}{E} \right)_{max}S \label{zzz}
\end{equation}
In (\ref{eremax}) - (\ref{costhetamax}) the $Q'$ terms are all squared and as a result, the sign of $Q'$ in (\ref{y}) is undetermined. To lift this degeneracy one uses (\ref{tanphie}) (\ref{x}) and (\ref{y}), obtaining
\begin{align}
 \tan\phi_{max} &= \frac{Q'}{P'} \notag\\
&= - \epsilon_3 \tan\phi_{max}, \label{ep3}
\end{align}
and it is clear from (\ref{ep3}) that $\epsilon_3=-1$. Hence the final equations for $S'$, $P'$ and $Q'$ are given by
 \begin{align}
 S' &= -\cos\theta_{max} \left(\frac{E'}{E}\right)_{max} S \label{S'}\\
 P'&= -\cos\phi_{max} \sin\theta_{max} \left(\frac{E'}{E}\right)_{max} S \label{P'}\\
 Q'&= -\sin\phi_{max} \sin\theta_{max} \left(\frac{E'}{E}\right)_{max} S \label{Q'}
 \end{align}
Expressions for the three arbitrary functions,  $S(r)$ $P(r)$ and $Q(r)$, are now easily obtained by integrating (\ref{S'}), (\ref{P'}) and (\ref{Q'}) over the radial coordinate. Since all the quantities to be integrated are specified in terms of $R_2$, it is sensible to integrate these expressions over the same radial coordinate. We find
\begin{align}
S(R_2) &=  S_0\exp\left[-\int^{R_2}_{R_{min}} \cos\theta_{max}\left(\frac{E'}{E}\right)_{max} dR_2^*\right], \hspace*{0.5cm} S(R_{min})=S_0 \label{SR} \\
P(R_2) &=P_0-\int^{R_2}_{R_{min}}\cos\phi_{max}\sin\theta_{max} \left(\frac{E'}{E}\right)_{max} S dR_2^* \label{PR}\\
Q(R_2)&=Q_0-\int^{R_2}_{R_{min}}\sin\phi_{max}\sin\theta_{max} \left(\frac{E'}{E}\right)_{max} S dR_2^* \label{QR}
\end{align}
where $P_0$ and $Q_0$ are initial values which can be set to zero. Since a constant rescaling of $S$ has no physical effect on the metric, one is free to choose the value of $S_0$. For simplicity, we will use $S_0=1$. Having obtained analytic expressions for the metric functions in terms of physical quantities, one can completely determine the Szekeres metric. While the quantities $\theta_{max}$, $\phi_{max}$ and $(E'/E)_{max}$ are not physical, they are related to the physical quantities $\theta_{\rho_{min}}$, $\phi_{\rho_{min}}$, $\rho_{min}$, $\rho_{LT}$ and $\rho_{AV}$, as was shown in \S \ref{solvingEre}.


\subsection{The deviation function and origin behaviour of \boldmath{$E'/E$}}
\label{originbehaviour}
As the origin is approached, by l'Hopital's rule $\rho_{AV}\rightarrow\rho_{LT}$, and we also expect $\rho_{min}\rightarrow\rho_{LT}$ since a pointlike dipole seems unphysical \footnote{Mathematically we merely require the terms $R'-RE'/E=R'(1-RE'/R'E)$ and $M'-3ME'/E=M'(1-3ME'/M'E)$ in (\ref{metric}) and (\ref{density}) to be well behaved at the origin, so $RE'/R'E \rightarrow 0$ is sufficient, since we already have $M \sim R^3$ there.}. Thus, it is apparent from (\ref{E'Emax}) that the origin value of $(E'/E)_{max}$ goes to $0/0$.  The conditions that ensure this quantity is finite at the origin are found by writing the density profiles, $\rho_{LT}$, $\rho_{min}$ and $\rho_{AV}$, as series expansions about $r=0$. On a constant time slice we make the co-ordinate choice $r=R$, and hence $R'=1$. The series expansions for $\rho_{LT}$ and $\rho_{min}$ are then given by
\begin{align}
 &\rho_{LT}=\rho_0 +\rho_1R + \rho_2R^2 + \rho_3R^3 + ... \label{rho_tl_series}\\
 &\rho_{min}=\zeta_0 + \zeta_1R + \zeta_2R^2 + \zeta_3R^3 + ...  \label{rho_min_series}
\end{align}
Rearranging (\ref{lt_density}) and inserting (\ref{rho_tl_series}) gives an expression for $M'$,
\begin{align}
M' &= \frac{\kappa}{2}[ \rho_0R^2 +\rho_1R^3 + \rho_2R^4 + \rho_3R^5 + ... ],
\end{align}
which can be integrated to yield an expression for $M$,
\begin{align}
M &= \frac{\kappa}{6} [ \rho_0R^3 +\frac{3}{4}\rho_1R^4 + \frac{3}{5}\rho_2R^5 + \frac{3}{6}\rho_3R^6 + ... ], \label{Mseries}
\end{align}
and substituting (\ref{Mseries}) into (\ref{rho_av}) gives
\begin{align}
\rho_{AV}  &=  \rho_0 +\frac{3}{4}\rho_1R + \frac{3}{5}\rho_2R^2 + \frac{3}{6}\rho_3R^3 + ... . \label{rho_av_series}
\end{align}
Now, inserting the series expansions (\ref{rho_tl_series}), (\ref{rho_min_series}) and (\ref{rho_av_series}) into the expression for $(E'/E)_{max}$ (\ref{E'Emax}) and taking the limit as $R \rightarrow 0$, one finds
\begin{align}
 \lim_{R\rightarrow0}\left(\frac{E'}{E}\right)_{max} &=\lim_{R\rightarrow0} \left| \left(\frac{1}{R} \right) \left[ \frac{(\zeta_0-\rho_0) + (\zeta_1-\rho_1)R + (\zeta_2-\rho_2)R^2 + (\zeta_3-\rho_3)R^3 + ...}{(\zeta_0 - \rho_0) + (\zeta_1-\frac{3}{4}\rho_1)R + (\zeta_2 - \frac{3}{5}\rho_2)R^2 + (\zeta_3 - \frac{3}{6}\rho_3)R^3 + ...}\right] \right| . \label{expansion1}
\end{align}
In order to avoid a divergence of (\ref{expansion1}) it is necessary to impose the condition
\begin{equation}
\zeta_0 = \rho_0. \label{condition1}
\end{equation}
This condition implies that no pointlike dipole can exist at $R=0$, if one requires a finite $E'/E$ value there. Whether a divergent $E'/E$ at the origin is physically acceptable or not is yet to be seen. However, since we are interested in numerical calculations, we will require a finite origin value for $E'/E$. Applying (\ref{condition1}) to (\ref{expansion1}), gives
\begin{align}
 \lim_{R\rightarrow0}\left(\frac{E'}{E}\right)_{max} &=\lim_{R\rightarrow0} \left| \left[ \frac{(\zeta_1-\rho_1) + (\zeta_2-\rho_2)R + (\zeta_3-\rho_3)R^2 + (\zeta_4-\rho_4)R^3 + ...}{(\zeta_1-\frac{3}{4}\rho_1)R + (\zeta_2 - \frac{3}{5}\rho_2)R^2 + (\zeta_3 - \frac{3}{6}\rho_3)R^3 + ...}\right] \right| . \label{expansion2}
\end{align}
In order to avoid a divergence of (\ref{expansion2}) it is necessary to impose the additional condition
\begin{equation}
\zeta_1 = \rho_1, \label{condition2}
\end{equation}
which gives
\begin{align}
 \lim_{R\rightarrow0}\left(\frac{E'}{E}\right)_{max} &=\lim_{R\rightarrow0} \left| \left[ \frac{(\zeta_2-\rho_2) + (\zeta_3-\rho_3)R + (\zeta_4-\rho_4)R^2 + (\zeta_5-\rho_5)R^3 + ...}{\frac{1}{4}\rho_1 + (\zeta_2 - \frac{3}{5}\rho_2)R + (\zeta_3 - \frac{3}{6}\rho_3)R + (\zeta_4 - \frac{3}{7}\rho_4)R^3 + ...}\right] \right| \label{expansion3} \\
 &= \left| 4 \left( \frac{\zeta_2 - \rho_2}{\rho_1}  \right) \right| \hspace{1cm} \rho_1 \neq 0.\label{limit1}
\end{align}
So with the conditions (\ref{condition1}), (\ref{condition2}) and $\rho_1 \neq 0$ satisfied, $E'/E$ will have a finite origin value given by (\ref{limit1}). However, (\ref{condition2}) also implies that such an arrangement would have a non-smooth central density profile, or `cusp', since $\rho_1 \neq 0$. In the case of smooth central density one has
\begin{align}
\rho_1 = 0, \label{condition_smooth}
\end{align}
 and so in order to avoid the divergence of (\ref{expansion3}), one must impose the further condition
 \begin{align}
 \zeta_2= \rho_2.	\label{condition3}
 \end{align}
Applying (\ref{condition_smooth}) and (\ref{condition3}) to (\ref{expansion3}) yields
\begin{align}
 \lim_{r\rightarrow0}\left(\frac{E'}{E}\right)_{max} &=\lim_{r\rightarrow0} \left| \left[ \frac{(\zeta_3-\rho_3) + (\zeta_4-\rho_4)R + (\zeta_5-\rho_5)R^2 + (\zeta_6-\rho_6)R^3 + ...}{\frac{2}{5}\rho_2 + (\zeta_3 - \frac{3}{6}\rho_3)R + (\zeta_4 - \frac{3}{7}\rho_4)R^3 + (\zeta_5 - \frac{3}{8}\rho_5)R^4 + ...}\right] \right| \label{expansion4} \\
 &= \left| \frac{5}{2} \left( \frac{\zeta_3 - \rho_3}{\rho_2} \right) \right| \hspace{1cm} \rho_2 \neq 0.\label{limit2}
\end{align}
So, if (\ref{condition1}), (\ref{condition2}), (\ref{condition_smooth}) (\ref{condition3}) and $\rho_2 \neq 0$ are all satisfied, the model has a smooth central density profile and the $E'/E$ value at the origin is finite, and given by (\ref{limit2}). Applying the same reasoning as above to higher orders, one finds that the limiting value, in general, is given by
\begin{align}
\lim_{r\rightarrow0}\left(\frac{E'}{E}\right)_{max} &=\left|  \frac{n+3}{n} \left( \frac{\zeta_{n+1} - \rho_{n+1}}{\rho_n} \right) \right| \hspace{1cm} \zeta_n = \rho_n \neq 0 \label{ErEgenlim}
\end{align}
where $n$ is the power in $R$ of the first non-zero term of $\rho_{LT}$ in (\ref{rho_tl_series}). This implies that for a finite $E'/E$ origin value, one requires $\rho_{min} = \rho_{LT}$ up to $n^{th}$ order in $R$. Also, for a non-zero origin limit, $\zeta_{n+1} \neq \rho_{n+1}$. In order to avoid having to choose $\rho_{min}$ in such a way, we instead define it in terms of $\rho_{LT}$ and a `deviation function', as follows.
\begin{align}
\rho_{min}(R) = \rho_{LT}(R) \left[ 1 - \mu(R) \right]	\label{deviation}
\end{align}
where $\mu$ satisfies
\begin{align}
0 \leq\mu(R) \leq 1 \hspace*{1cm} \mu(0) = 0 \label{mucond}
\end{align}
and
\begin{align}
\frac{d^{i} \mu}{d R^i} |_{R=0}=0  \hspace{1cm} i=1..n \label{mucond1}
\end{align}
Choosing any function $\mu$ in this way is sufficient to ensure that the origin value of $E'/E$ is finite. If, in addition to (\ref{mucond}) and (\ref{mucond1}), one has
\begin{align}
\frac{d^{n+1} \mu}{d R^{n+1}} \neq 0, \label{mucond2}
\end{align}
the origin value of $E'/E$ will be given by (\ref{ErEgenlim}). Clearly, if (\ref{mucond2}) is not satisfied, $E'/E$ will have an origin value of zero\footnote{It is worth noting that the origin value of $E'/E$ is a coordinate dependant quantity (since it contains a prime), and the treatment above is only valid for the coordinate choice $r=R$. In our original formulation we made the coordinate choice $r=M$, which turned out to be far more restrictive on the allowable density profiles which produce a finite origin value of $E'/E$.}.


\subsection{Expressing \boldmath{$\rho_{max}$} in terms of \boldmath{$\rho_{min}$} and \boldmath{$\rho_{LT}$}}
It will be useful to have an expression for the maximum density profile, $\rho_{max}$, in terms of the density profiles, $\rho_{min}$ and $\rho_{LT}$,  that does not involve $E'/E$. Rearranging (\ref{ere}) such that $\rho$ is the subject, and separating the right hand side into two terms, one finds 
\begin{align}
\rho = \rho_{LT} + \frac{\frac{E'}{E} ( \rho_{LT} - \rho_{AV})}{\frac{R'}{R} - \frac{E'}{E}}
\label{rho=lt+}
\end{align}
Now, evaluating (\ref{rho=lt+}) at the density maximum and exploiting the anti-symmetrical property of $E'/E$ (\ref{antisym}), leads to
\begin{align}
\rho_{max} &= \rho_{LT} - \frac{\frac{E'}{E}|_{\rho_{min}} ( \rho_{LT} - \rho_{AV})}{\frac{R'}{R} + \frac{E'}{E}|_{\rho_{min}}} 
\end{align}
Substituting for $E'/E|_{\rho_{min}}$ from (\ref{ere}) with $\rho=\rho_{min}$,  and simplifying, one finds
\begin{align}
\rho_{max} &= \rho_{LT} + \frac{(\rho_{LT} - \rho_{min})(\rho_{LT} - \rho_{AV} )}{(\rho_{AV} - \rho_{min}) + (\rho_{LT} - \rho_{min})}.
\label{rho_max}
\end{align}
So, by specifying the minimum and LT density profiles, $\rho_{min}(r)$ and $\rho_{LT}(r)$, one fixes the maximum density profile, $\rho_{max}(r)$.

\subsection{Evolution considerations}
\subsubsection{LT density}
The LT-density, $\rho_{LT}$, will be specified on the initial and final time slices, as was done in \cite{Krasinski:2001yi}. When considering the time evolution of the model, the value of $\rho_{LT}$ on intermediate time slices is calculated according to (\ref{rholta}) and the evolution of $a$ at the origin found from the applicable one of (\ref{RevolE}) (\ref{RevolP}) and (\ref{RevolH}).

\subsubsection{Reconstructing \boldmath{$\rho_{min}$} from \boldmath{$(E'/E)_{max}$} during evolution}
When reconstructing the model evolution it will be convenient to have an expression for $\rho_{min}$ in terms of $(E'/E)_{max}$, instead of being in terms of $(E'/E)_{\rho_{min}}$. This is because $(E'/E)_{max}$ is constant in time. From the analysis of \S \ref{solvingEre}, we have
\begin{align}
\left(\frac{E'}{E}\right)_{\rho_{min}} = \chi \left( \frac{E'}{E} \right)_{max} 
\label{ererhomin}
\end{align}
where
\begin{align}
\chi \equiv - {\rm sign} [R'(\rho_{LT}-\rho_{AV})] 
\label{chi}
\end{align}
Hence, setting $\rho=\rho_{min}$ in (\ref{rho=lt+}), and substituting in (\ref{ererhomin}),  gives
\begin{align}
\rho_{min} = \rho_{LT} + \frac{\chi \left( \frac{E'}{E} \right)_{max}  ( \rho_{LT} - \rho_{AV})}{\frac{R'}{R} - \chi \left( \frac{E'}{E} \right)_{max} }.
\label{rhominevol}
\end{align}
Similarly, for $\rho_{max}$, one can write
\begin{align}
\rho_{max} = \rho_{LT} - \frac{\chi \left( \frac{E'}{E} \right)_{max}  ( \rho_{LT} - \rho_{AV})}{\frac{R'}{R} + \chi \left( \frac{E'}{E} \right)_{max} }.
\label{rhomaxevol}
\end{align}
Therefore, having calculated the profiles $(E'/E)_{max}(M)$, $R(t,M)$, $R'(t,M)$ and $\rho_{LT}(t,M)$ one may reconstruct the evolution of $\rho_{min}(t,M)$ and $\rho_{max}(t,M)$.

\subsubsection{Approximating \boldmath{$R_1/R_2$}}
\label{R1/R2sec}
Since we choose to specify, at the final time, the density profiles and dipole parameters as a function of $R_2$, it will be useful to know the approximate range of $R_1$ which contains the same mass, prior to integrating the final density profile. This will allow one to specify the fluctuation length scale of the initial LT density profile, relative to the same set of worldlines. Thus, an approximate expression for $R_1/R_2$ is needed. Rearranging (\ref{rho_av}) and evaluating at the initial and final time, and dividing the two, leads to
\begin{align}
R = \left( \frac{6M}{\rho_{AV}} \right)^{1/3} \Rightarrow \frac{R_1}{R_2} = \left( \frac{\rho_{AV,2}}{\rho_{AV,1}} \right)^{1/3}.
\label{R1R2}
\end{align}
Assuming the simulated density profiles are asymptotically uniform at large $R$, approaching some constant value, say $\rho_{BG}$, and if all the fluctuations are contained within the simulation, then we can make the rather crude approximation at the boundary of the simulation
\begin{align}
\rho_{AV} \approx \rho_{BG},
\end{align}
and thus (\ref{R1R2}) gives
\begin{align}
\frac{R_1}{R_2} \approx \left( \frac{\rho_{BG,2}}{\rho_{BG,1}} \right)^{1/3}.
\label{R1R2approx}
\end{align}
Alternatively, recalling that for a flat FLRW evolution during the radiation dominated era the scale factor grows like $a \propto t^{2/3}$. If we assume that our simulation is not too different from parabolic evolution at the extremity of the simulation then we can approximate the growth rate of the areal radius to that of the scale factor from above. Hence, one can write
\begin{align}
\frac{R_1}{R_2} \approx \left( \frac{t_1}{t_2} \right)^{2/3}.
\label{R1R2approx2}
\end{align}
So, either (\ref{R1R2approx}) or (\ref{R1R2approx2}) can be used to approximate the range of $R_1$, given $R_2$.

\subsection{The algorithm}
\label{algorithm}
The above analysis has equipped us with the tools to construct the Szekeres model that evolves between two time slices, given initial and final density data. The procedure is broken into two parts. Firstly, obtaining the six arbitrary functions which define the metric from the specified initial and final data. And secondly, calculating the model evolution from those arbitrary functions. To begin, one must specify the following quantities:
\begin{itemize}
 \item $\rho_{LT,1}(R_1)$: The LT-density profile at $t = t_1$
 \item $\rho_{LT,2}(R_2)$: The LT-density profile at $t = t_2$
 \item $\mu(R_2)$: The `deviation function' at $t=t_2$
 \item $\theta_{\rho_{min}}(R_2)$: The density minimum orientation angle, $\theta$, at $t=t_2$
\item $\phi_{\rho_{min}}(R_2)$: The density minimum orientation angle, $\phi$, at $t=t_2$
\end{itemize}
The approximate range of $\rho_{LT,1}$ is related to that of $\rho_{LT,2}$ by (\ref{R1R2approx}) or (\ref{R1R2approx2}). Now, the metric functions $M(R_2)$, $f(R_2)$, $t_B(R_2)$, $S(R_2)$, $P(R_2)$ and $Q(R_2)$ are obtained as follows: 

\bigskip\noindent Evaluate the integral (\ref{M(R)}) (with $\rho=\rho_{LT}$) at the initial and final times, to find $M(R_1)$ and $M(R_2)$. Interpolation then gives $R_1(M)$ and $R_2(M)$, for a set of  worldlines. Now, for each particle worldline (at each $M$ value):
\begin{itemize}
\item Calculate  $a_1(M)$ and $a_2(M)$ using (\ref{ax}). In the case where $M_{min}=0$, the origin value, $a(0)$, is given by (\ref{ai0}) (with $\rho=\rho_{LT}$).
\item Evaluate the inequalities (\ref{ineq1}) (\ref{ineq2}),(\ref{ineq3}),(\ref{ineq4}) (\ref{ineq5}), and hence determine the evolution type of that worldline.
\item Determine the value of $x$ which solves the relevant $\psi(x)=0$ equation, (\ref{psix1}) (\ref{psix2}) (\ref{psix3}) (\ref{psix4}) (\ref{psix5}). A bisection method is good for this - refer to Table \ref{bisect} for the initial rage over which to bisect.
\item Calculate $f(M)$ using the one of (\ref{f1}) (\ref{f2}) (\ref{f3}) (\ref{f4}) (\ref{f5}) which is applicable to the evolution type.
\item Calculate  $t_b(M)$ using the one of (\ref{tb1}) (\ref{tb2}) (\ref{tb3}) (\ref{tb4}) (\ref{tb5}) which is applicable to the evolution type.


\item Calculate $E'/E|_{max} (M)$ on the final time slice using (\ref{E'Emax}) with (\ref{deviation}) and (\ref{rho_av}). At the origin $\rho_{AV}(0) = \rho_{LT}(0)$, and $E'/E|_{max} (0)$ is given by (\ref{ErEgenlim}) if (\ref{mucond2}) is satisfied. If (\ref{mucond2}) is not satisfied, the origin value of $E'/E|_{max}$ is zero.


\item Determine the sign of $R'(\rho_{LT}-\rho_{AV})$ at final time, and hence calculate $\sin\theta_{max}$, $\sin\phi_{max}$, $\cos\theta_{max}$ and $\cos\phi_{max}$ using either (\ref{cs1}) or (\ref{cs2}).

\item Calculate $S$, $P$ and $Q$ using (\ref{SR}), (\ref{PR}) and (\ref{QR}).

\end{itemize}
Having determined the metric functions one can now check for regular origins, spatial maxima or minima and shell crossings (see \S6.1 in \cite{Hellaby:2002nx}). Going on to reconstruct the model evolution, on a set of intermediate time slices, one calculates
\begin{itemize}
\item $a(M)$ using the one of (\ref{RevolE}) (\ref{RevolP}) (\ref{RevolH}) which is applicable to the evolution type.
\item $\rho_{LT}(M)$ using (\ref{rholta}).
\item $R(M)$ using (\ref{ax}).
\item $R'(M)$ using (\ref{RMprime}).
\item $\rho_{AV}(M)$  using (\ref{rho_av})
\item $\chi(M)$ using (\ref{chi}).
\item $\rho_{min}(M)$ and $\rho_{max}(M)$ using (\ref{rhominevol}) and (\ref{rhomaxevol}).
\end{itemize}
\section{Numerical simulations}
\label{numerics}

\subsection{Implementation}
\label{implementation}
A script was written in MATLAB to implement the algorithm outlined in \S \ref{algorithm}. 
Some practical aspects of the implementation are outlined below.
\begin{itemize}
 \item Evaluation of the integral in (\ref{M(R)}), to obtain $M(R_1)$ and $M(R_2)$, is computed using the adaptive Gauss-Kronrod quadrature package, {\it quadgk}. The inverses, $R_1(M)$ and $R_2(M)$, defined over approximately the same mass rage, are then found. The function {\it interp1} is used to interpolate the values of $R_1$ at the same $M$ values for which $R_2$ is defined.
\item The quantities $a_1$ and $\rho_{AV,1}$ are best calculated using the values of $R$ and $M$ given by the integration of (\ref{M(R)}), prior to interpolation. Once $a_1$ and $\rho_{AV,1}$ are known, the function {\it interp1} is the used to interpolate values corresponding to the same $M$ values for which $R_2$ is defined.
\item The function m-file {\it solve\_phi\_x.m} was written to solve the various $\psi(x)=0$ equations by the bisection method.
\item Evaluation of the integral in (\ref{SR}) is computed using the ODE solver {\it ode45}, while the integrals in (\ref{PR}) and (\ref{QR}) are computed using the adaptive Simpson quadrature package {\it quad}. The function m-file {\it myinerpfun.m} was written so that the intergrands in (\ref{SR}) (\ref{PR}) and (\ref{QR}) are continuously defined, and thus amenable to integration with {\it ode45} and {\it quad}.
\item For non-parabolic world lines, the evolution of $R$ is not given explicitly in terms of $t$, but instead it is parametrized by $\eta$. This necessitates the interpolation of $R$ on constant time slices. Along constant-$M$ slices an $N\times1$ vector of $\eta$-values is created ranging from $\eta_1$ to $\eta_2$ \footnote{For regions that are recollapsing at $t=t_2$, the final phase is given by $\eta_2 = 2\pi - \arccos(1-a_2x)$}, allowing $R(\eta)$ and the corresponding $t(\eta)$ to be calculated.  A spline cure is fitted to $R(t)$ at each $M$ value using {\it spline}, allowing interpolation of $R$ at constant time slices.
\item Derivatives, such as $R'(t,M)$, $\dot R(t,M)$ and $a_{,M}(t,M)$, are computed using the package {\it gradient}, which uses a 3-point finite difference method.
\end{itemize}

\subsection{Model profiles}
\label{model_profiles}
Since the aims of this paper are to develop a model construction procedure and implement it in software, investigating elaborate models using this software is thus outside our scope. Our choice of model profiles was therefore predominantly motivated by convenience, the intention being to sufficiently demonstrate the functioning of the software. We use geometric units, in which $c=1=G$ and fix the scale freedom of GR by considering $10^{15} M_{\odot}$ to be one mass unit, as in \cite{Krasinski:2001yi}. The corresponding geometric length and time units are related by
\begin{align}
1 M_G &= 10^{15} \: M_{\odot}, \notag\\
\Rightarrow 1 L_G &= M_G \frac{G}{c^2} = 48 \: \rm{pc}, \notag\\
\Rightarrow 1 T_G  &= M_G \frac{G}{c^3} = 156 \: \rm{yr}.
\end{align}
In all of the simulations we choose the final time to be, approximately, the current age of the Universe, and the initial time to be, approximately, the time of recombination.  That is
\begin{align}
&t_1 = 100 \: \rm{k \: yr} = 641 \: T_G, \\
&t_2 = 10 \: \rm{G \:yr} = 6.41 \times 10^7 \: T_G.
\end{align}
Similarly, background densities are given by
\begin{align}
\rho_{BG,1} = 8 \times 10^{-17} \: \rm{kg/m^3} = 1.3 \times 10^{-7} \: M_G/L_G^3, \notag\\
\rho_{BG,2} = 8 \times 10^{-27} \: \rm{kg/m^3} = 1.3 \times 10^{-17} \: M_G/L_G^3. \label{run1_bg}
\end{align}
In all cases, the range of $R_2$ over which we simulate the chosen profiles is $10^{6} \: L_G$, which is equivalent to $48$ Mpc. 

As developed above, the models will be specified in the following way.  On the initial time slice, the ``LT-density'' is given.  This shows the variation of the ``equatorial" density (the density where $E'/E = 0$) with areal radius at that time.  A similar profile is given for the final time.  The strength of the density dipole is given at the final time via (\ref{deviation}), as a deviation from the ``LT-density'', while two more profiles specify the $\theta$ \& $\phi$ orientation of this minimum.  Thus as a rough guide, the locus of shell-minimum density follows $(R_2,\theta_{\rho_{min}},\phi_{\rho_{min}})$.\footnote{A detailed visualisation would require a careful treatment of the non-concentricity of the shells, and would be complicated by the fact that the space is curved in general, while any plot would be in flat space.}

\subsubsection{Run \#1}
We first investigate the simple case where the dipole orientation is unchanging and $E'/E|_{max}(0)=0$, and thus choose the `deviation function' such that it does not satisfy (\ref{mucond2}). The chosen profiles are
\begin{align}
\rho_{LT,1}(R_1) &= \rho_{BG,1} \left( \frac{1.00003 +(8 \times 10^{-5}) R^3}{1+ (8 \times 10^{-5}) R^3}\right) \notag\\
\rho_{LT,2}(R_2) &=  \rho_{BG,2} \left(  \frac{10^5 + 10^{-10} R^2 + 10^{-17}R^3 }{10 + 10^{-10} R^2 + 10^{-17}R^3} \right)\notag\\
\mu(R_2) &= 10^{-23} R^4  \exp  \left( - (4 \times 10^{-24})R^4 \right) \notag\\
\theta_{\rho_{min}}(R_2) &= \frac{\pi}{4} \notag\\
\phi_{\rho_{min}}(R_2) &= \frac{\pi}{4},
\end{align}

\subsubsection{Run \#2}
Next, we investigate the case of $R$-dependant dipole angles and where $E'/E|_{max}(0) \neq 0$, and thus choose the `deviation function' such that it does satisfy (\ref{mucond2}). The chosen profiles are
\begin{align}
\rho_{LT,1}(R_1) &= \rho_{BG,1} \left( \frac{1.00003 + (10^{-5}) R^3}{1+ (10^{-5}) R^3}\right) \notag\\
\rho_{LT,2}(R_2) &=  \rho_{BG,2} \left(  \frac{(5 \times 10^4) + 10^{-10} R^2 + 10^{-17}R^3 }{10 + 10^{-10} R^2 + 10^{-17}R^3} \right)\notag\\
\mu(R_2) &= 10^{-17} R^3  \exp  \left( - (4 \times 10^{-18})R^3 \right) \notag\\
\theta_{\rho_{min}}(R_2) &= \frac{\pi}{4} + (\frac{\pi}{2} \times 10^{-6} ) R \notag\\
\phi_{\rho_{min}}(R_2) &=    (2 \pi \times 10^{-6})  R,
\end{align}

\subsubsection{Run \#3}
Though the profiles are superficially similar to Run \#2, this case has a strongly varying late time $\rho_{max}$ profile. The chosen profiles are
\begin{align}
\rho_{LT,1}(R_1) &= \rho_{BG,1} \left( \frac{1.00003 +(1 \times 10^{-5}) R^3}{1+ (1 \times 10^{-5}) R^3}\right) \notag\\
\rho_{LT,2}(R_2) &=  \rho_{BG,2} \left(  \frac{1 + (2\times 10^{-10}) R^2 + (2.9 \times 10^{-17})R^3 }{10 + (2 \times 10^{-10}) R^2 + (2.9 \times 10^{-17})R^3 } \right)\notag\\
\mu(R_2) &= (2.2 \times 10^{-18}) R^3  \exp  \left( - (1.2 \times 10^{-17})R^3 \right) \notag\\
\theta_{\rho_{min}}(R_2) &= \frac{\pi}{4} + (\frac{\pi}{2} \times 10^{-6} ) R \notag\\
\phi_{\rho_{min}}(R_2) &=    (2 \pi \times 10^{-6})  R,
\end{align}

\subsection{Results and discussion}
\label{results_discussion}
\subsubsection{Run \#1}

\begin{figure}
\begin{center}
\includegraphics[scale=0.8]{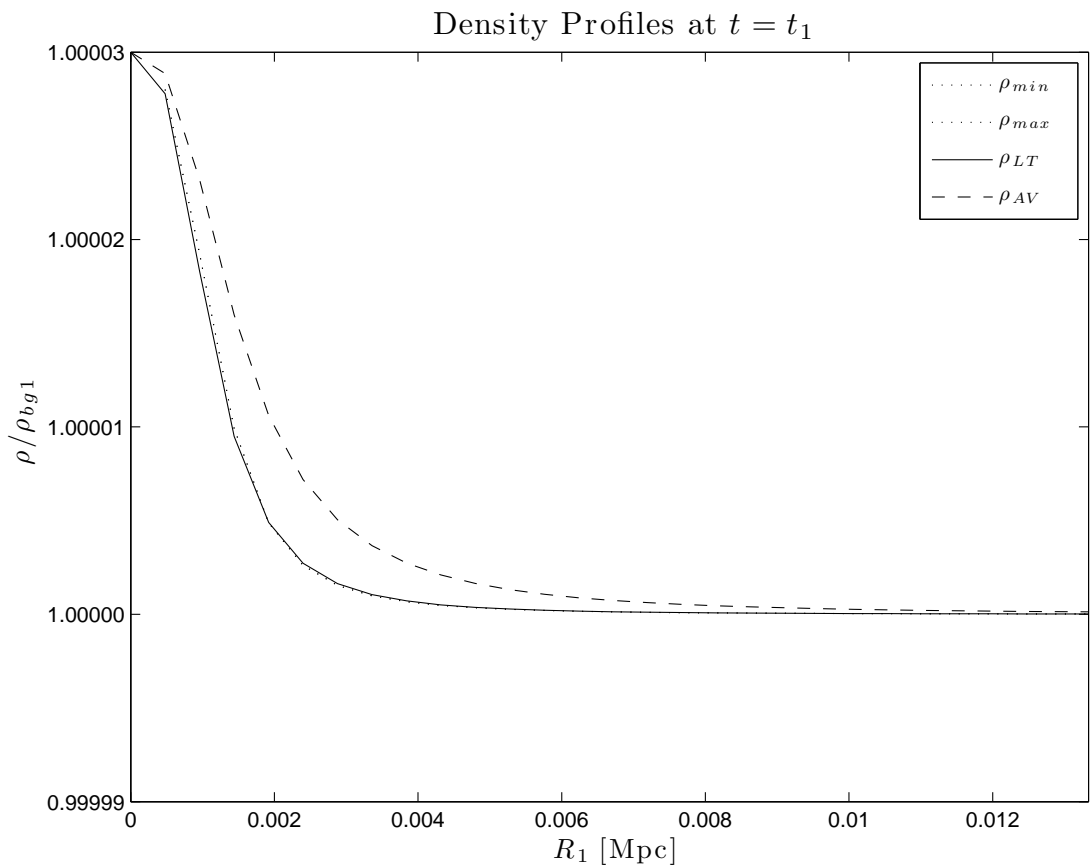}
\end{center}
\caption[Initial Density Profiles from Run \#1]{\textbf{Initial Density Profiles from Run \#1} - The minimum, maximum, internal average and LT density profiles at $t=t_1$ from Run \#1. The LT-density was specified and the rest were calculated. (The profiles extend out to $R_1=0.27$) }
\label{run1_rho1}
\end{figure}

\begin{figure}
\begin{center}
\includegraphics[scale=0.6]{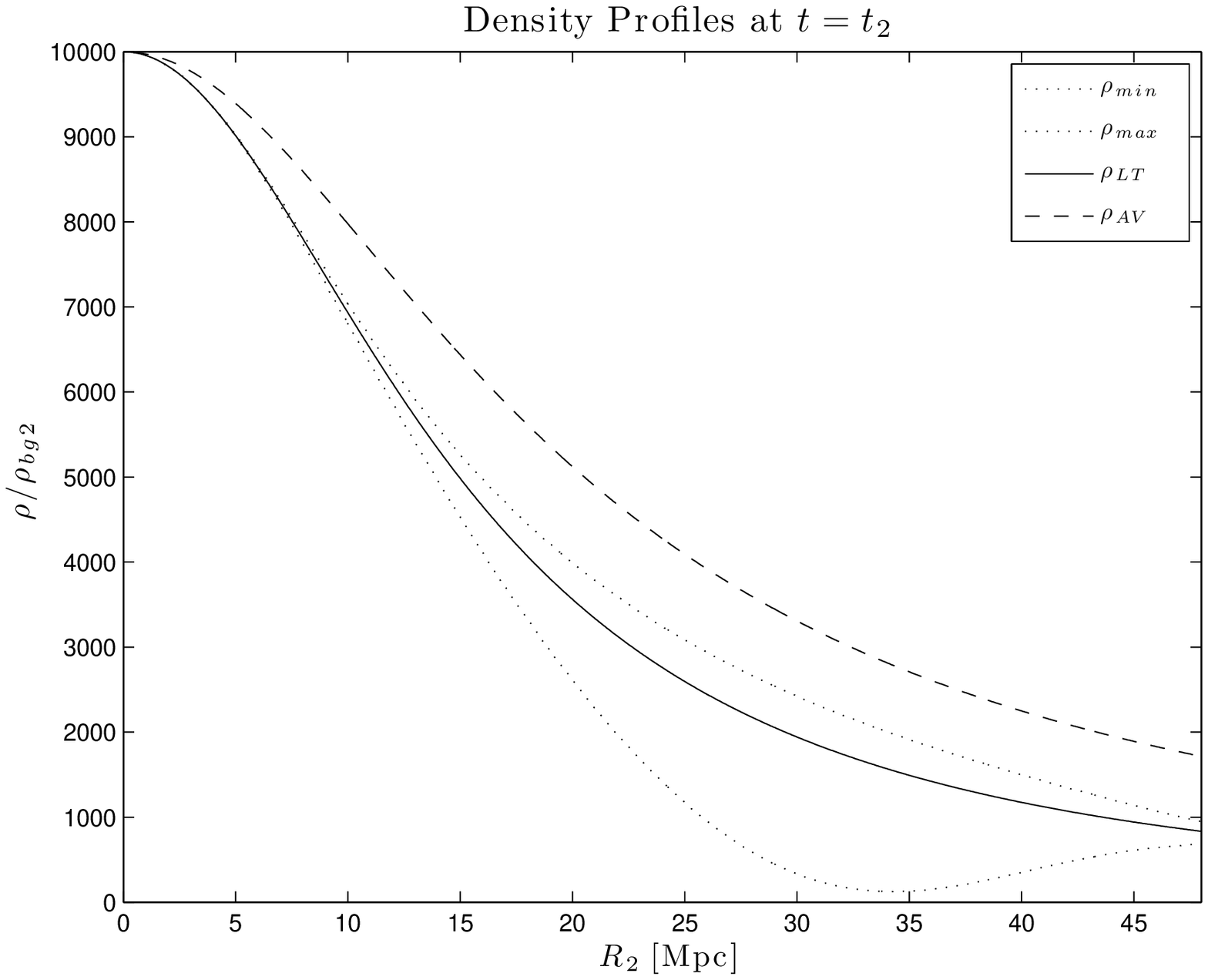}
\end{center}
\caption[Final Density Profiles from Run \#1]{\textbf{Final Density Profiles from Run \#1} - The minimum, maximum, internal average and LT density profiles at  $t=t_2$ from Run \#1. The minimum and LT-density were specified. The rest were calculated.}
\label{run1_rho2}
\end{figure}

\begin{figure}
\begin{center}
\includegraphics[scale=0.6]{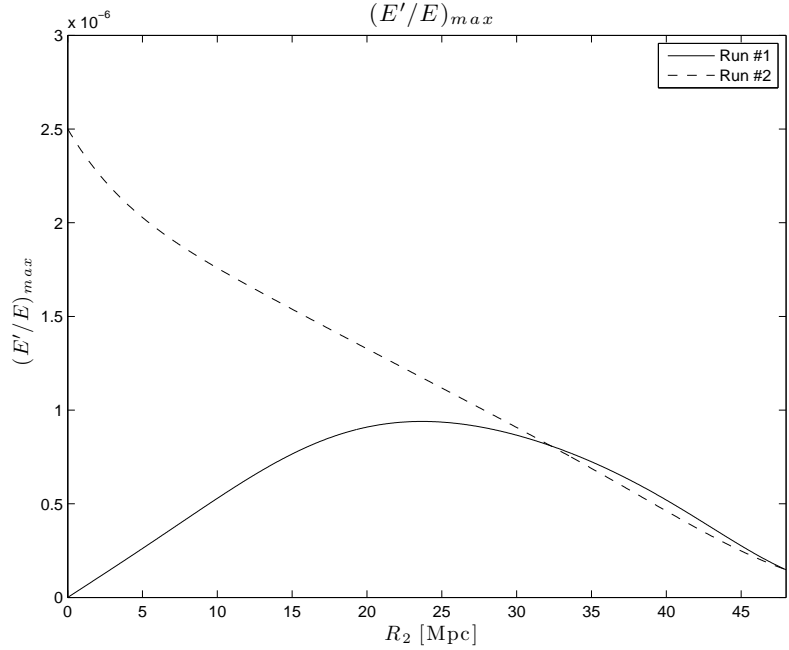}
\end{center}
\caption[The Function $E'/E|_{max}$ from Run \#1 and Run \#2]{\textbf{The Function $E'/E|_{max}$ from Run \#1 and Run \#2} - The dipole function $E'/E|_{max}$ from Run \#1 (solid line) and Run \#2 (dashed line).}
\label{run1_ere}
\end{figure}

\begin{figure}
\begin{center}
\includegraphics[scale=0.8]{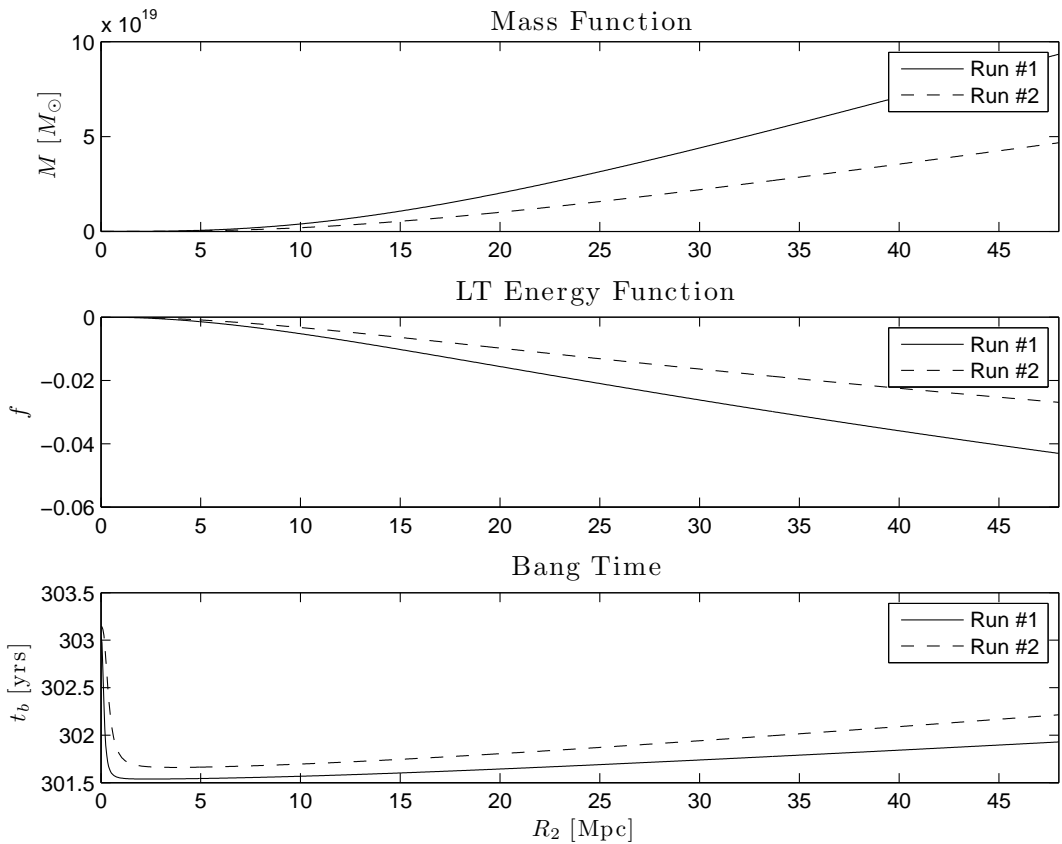}
\end{center}
\caption[Arbitrary Functions $M$, $f$ and $t_b$ from Run \#1 and Run \#2]{\textbf{Arbitrary Functions $M$, $f$ and $t_b$ from Run \#1 and Run \#2} - The arbitrary functions which are common to both Szekeres and LT modes; $M$, $f$ and $t_b$, from Run \#1 (solid line) and Run \#2 (broken line)}
\label{run1_mftb}
\end{figure}

\begin{figure}
\begin{center}
\includegraphics[scale=0.8]{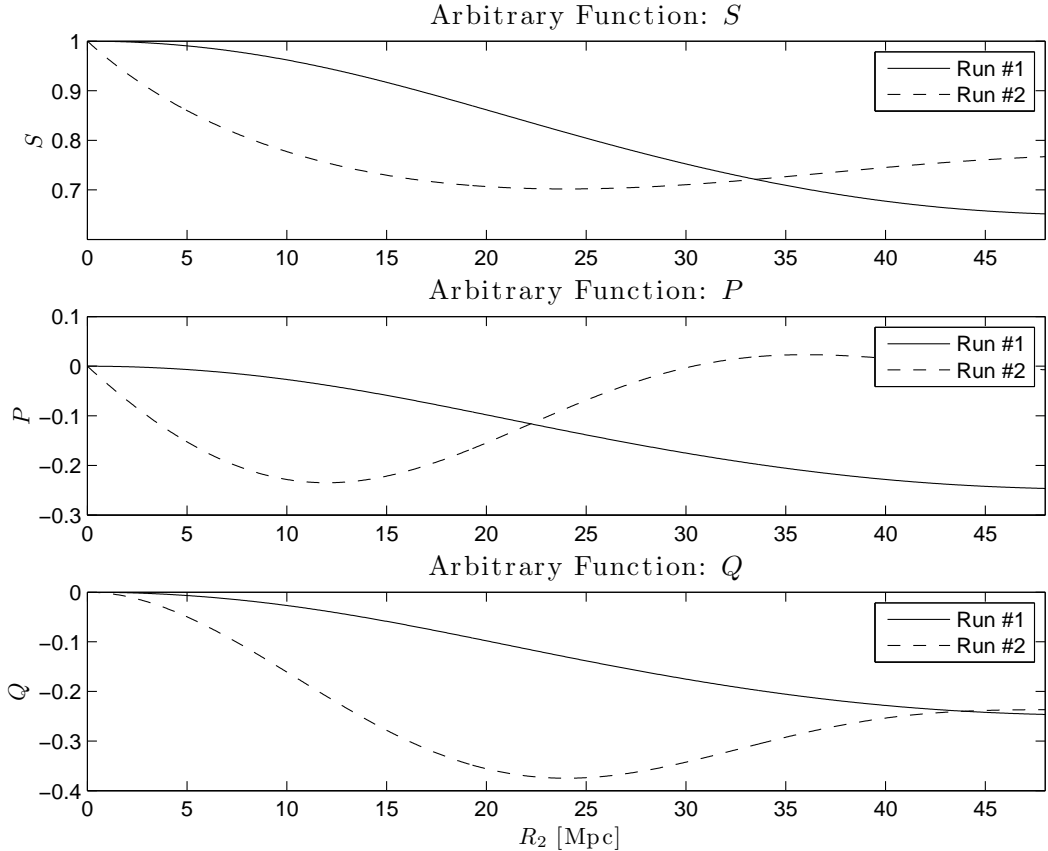}
\end{center}
\caption[Arbitrary Functions $S$, $P$ and $Q$ from Run \#1]{\textbf{Arbitrary Functions $S$, $P$ and $Q$ from Run \#1  and Run \#2} - The arbitrary functions which are unique to Szekeres models; $S$, $P$ and $Q$, from Run \#1 (solid line) and Run \#2 (broken line)}
\label{run1_spq}
\end{figure}

\begin{figure}
\begin{center}
\includegraphics[scale=0.6]{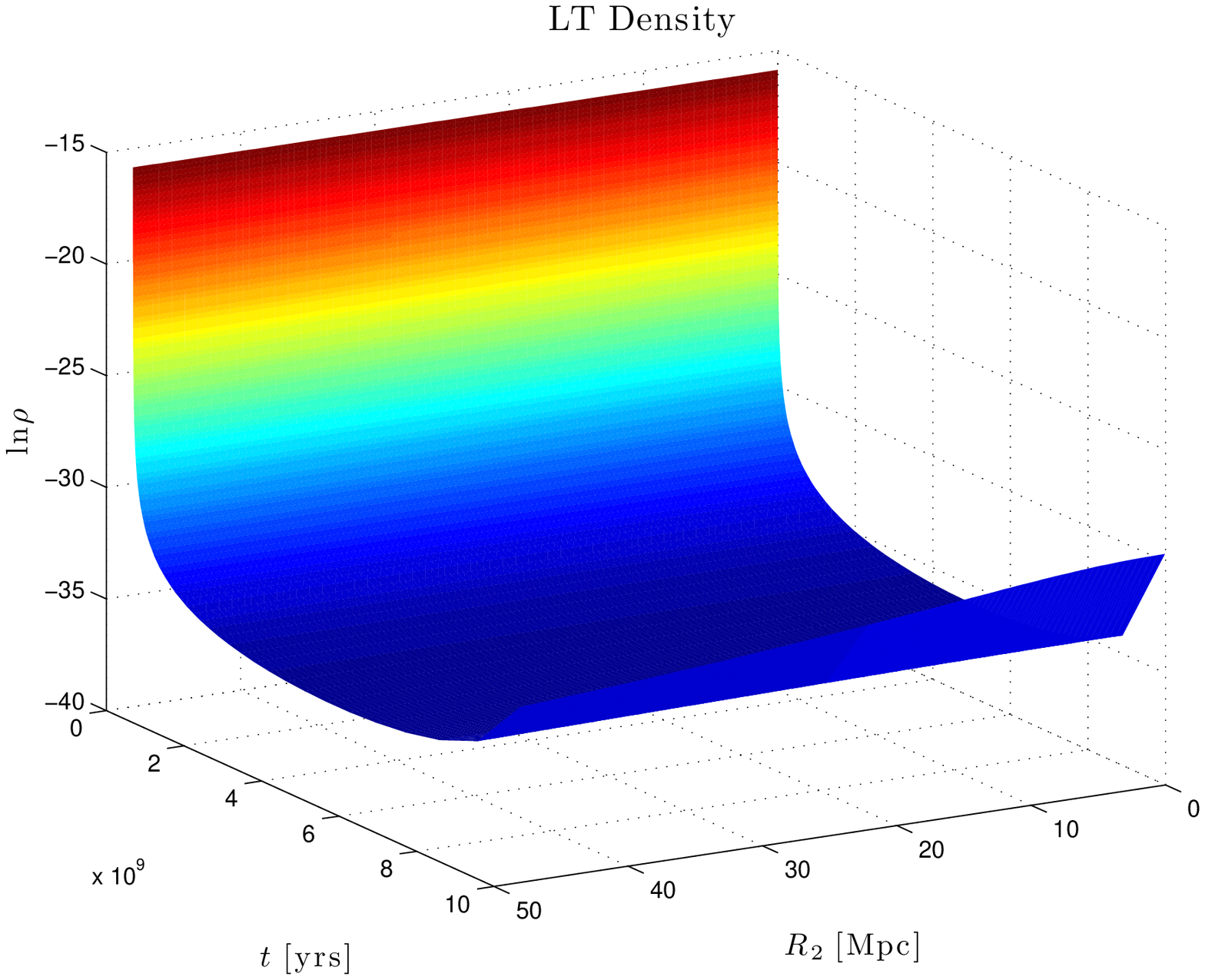}
\end{center}
\caption[Evolution of the LT-density from Run \#1]{\textbf{Evolution of the LT-density from Run \#1} - The evolution of the LT-density profile, $\rho_{LT}$. Log scale}
\label{run1_rholt}
\end{figure}

\begin{figure}
\begin{center}
\includegraphics[scale=0.6]{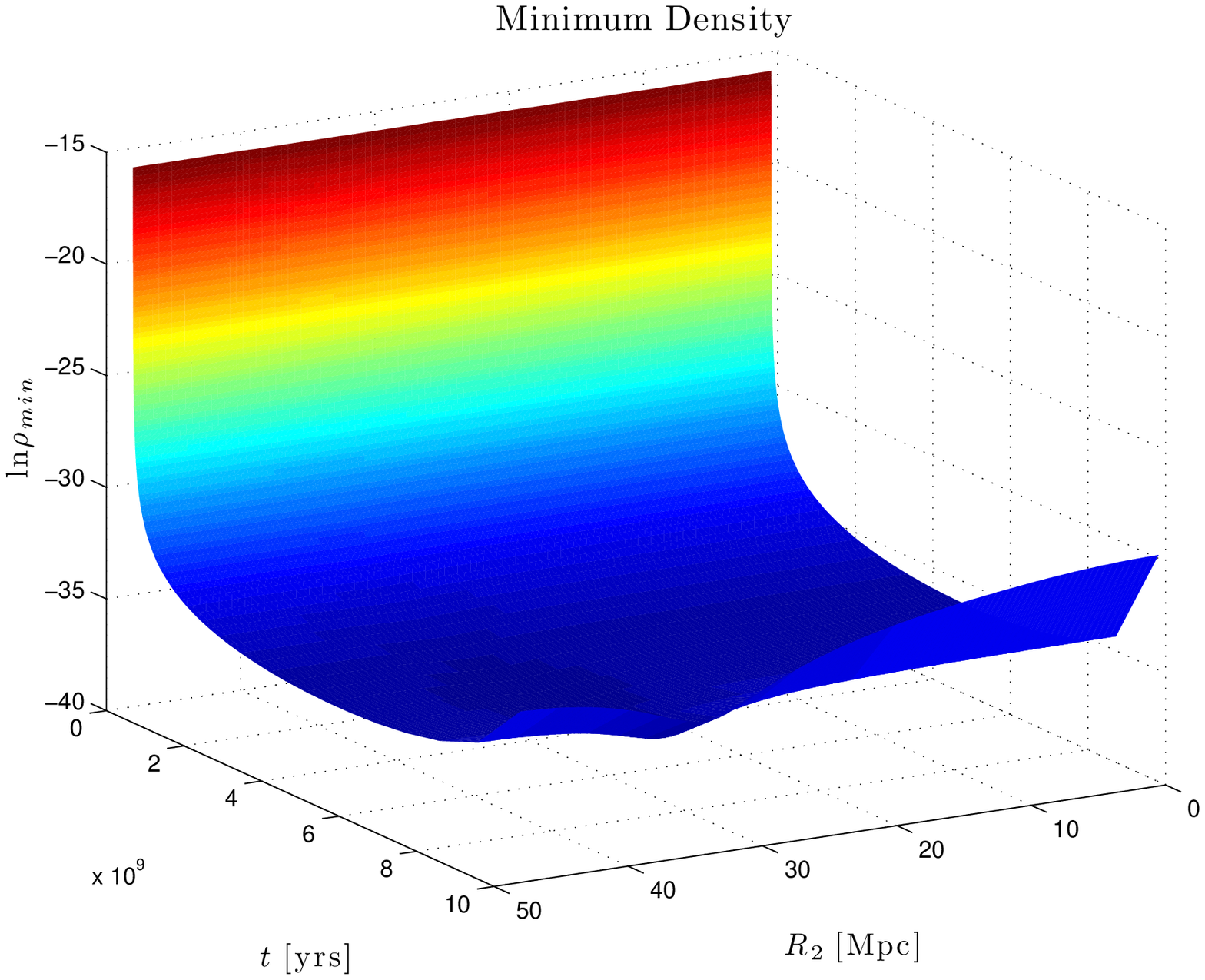}
\end{center}
\caption[Density Dipole Minimum from Run \#1]{\textbf{Density Dipole Minimum from Run \#1} - The evolution of the density dipole minimum, $\rho_{min}$. Log scale}
\label{run1_rhomin}
\end{figure}

\begin{figure}
\begin{center}
\includegraphics[scale=0.6]{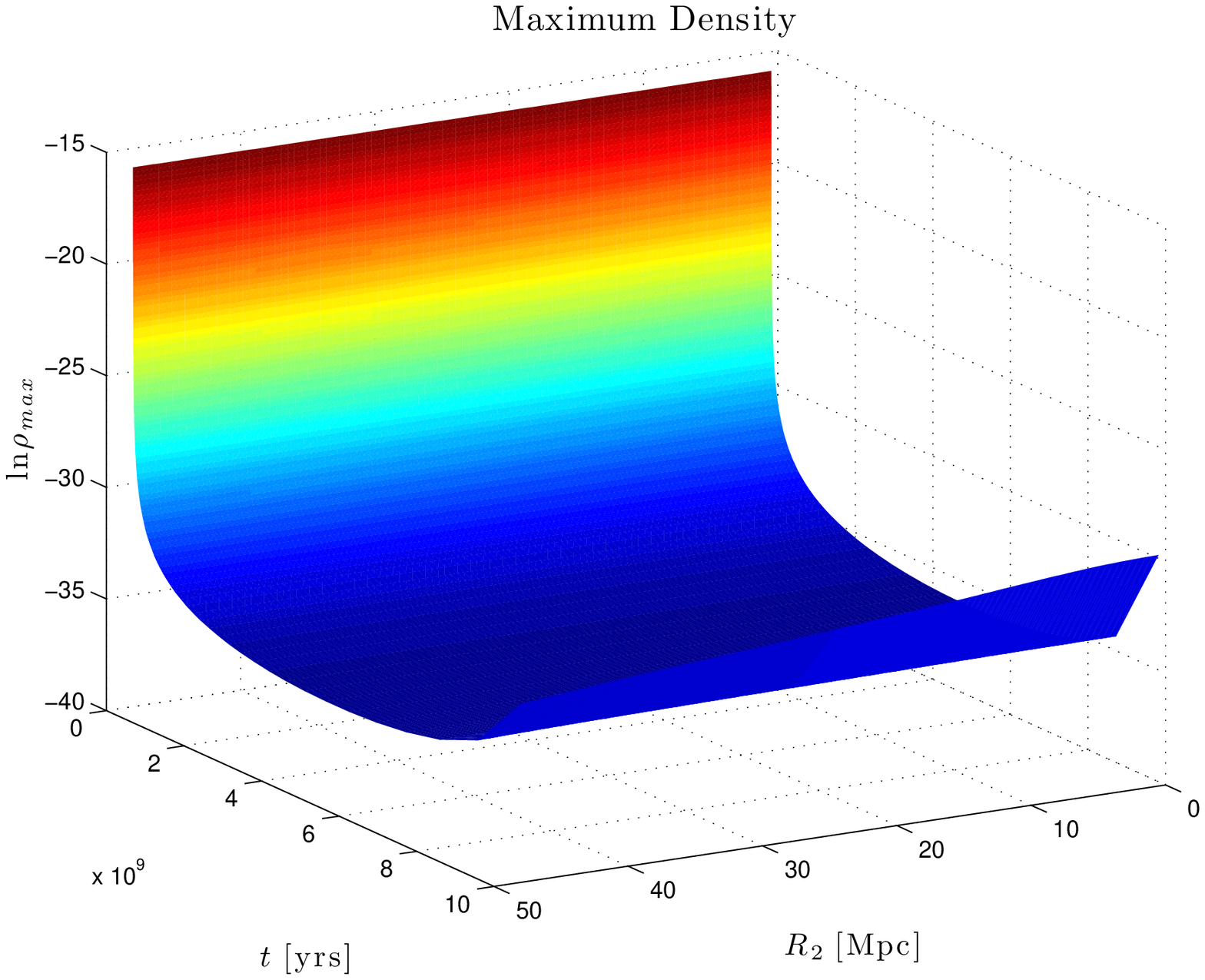}
\end{center}
\caption[Density Dipole Maximum from Run \#1]{\textbf{Density Dipole Maximum from Run \#1} - The evolution of the density dipole maximum, $\rho_{max}$. Log scale}
\label{run1_rhomax}
\end{figure}

The density profiles $\rho_{LT}$, $\rho_{min}$, $\rho_{max}$ and $\rho_{AV}$ at the initial and final times are shown in Figures \ref{run1_rho1} and \ref{run1_rho2}, respectively. Both have an over-density at the origin, resulting in $\rho_{AV} > \rho_{LT}$ for all $(t,r)$, and thus the relationship between $E'/E|_{max}$ and $E'/E|_{\rho_{min}}$ is constant.  At the final time, the large deviation in $\rho_{min}$ away from $\rho_{LT}$ (which was specified via the function $\mu$) causes little deviation in $\rho_{max}$ away from $\rho_{LT}$. The large deviation in $\rho_{min}$ at the final time is sourced by a much smaller deviation from the initial time. The $(\theta,\phi)$ orientation of the density dipole is constant. The function $E'/E|_{max}$ shown in Figure \ref{run1_ere} has an origin value of zero, as expected from our choice of $\mu$. Figure \ref{run1_mftb} shows the LT arbitrary functions, $M$, $f$ and $t_b$. Since there are no regions where $\rho_{LT}=0$, we see that $M$ in the top pane of Figure \ref{run1_mftb} is always increasing. The LT energy function in the centre pane has $f<0$, and thus all worldlines are elliptical. The bottom pane shows the bang time, which is decreasing close to the origin and then becomes increasing further away. This will produce shell crossings, however, given the gradual slope we expect these to occur very soon after the bang, well before $t_1$, when we anticipate our model will not be valid anyway, due to its dust equation of state. The Szekeres arbitrary functions, $S$, $P$ and $Q$, are shown in Figure \ref{run1_spq}. Surfaces showing the full time-evolution of $\rho_{LT}$, $\rho_{min}$ and $\rho_{max}$ are shown in Figures \ref{run1_rholt} \ref{run1_rhomin} and \ref{run1_rhomax}, respectively.

%
%

\subsubsection{Run \#2}
\begin{figure}
\begin{center}
\includegraphics[scale=0.6]{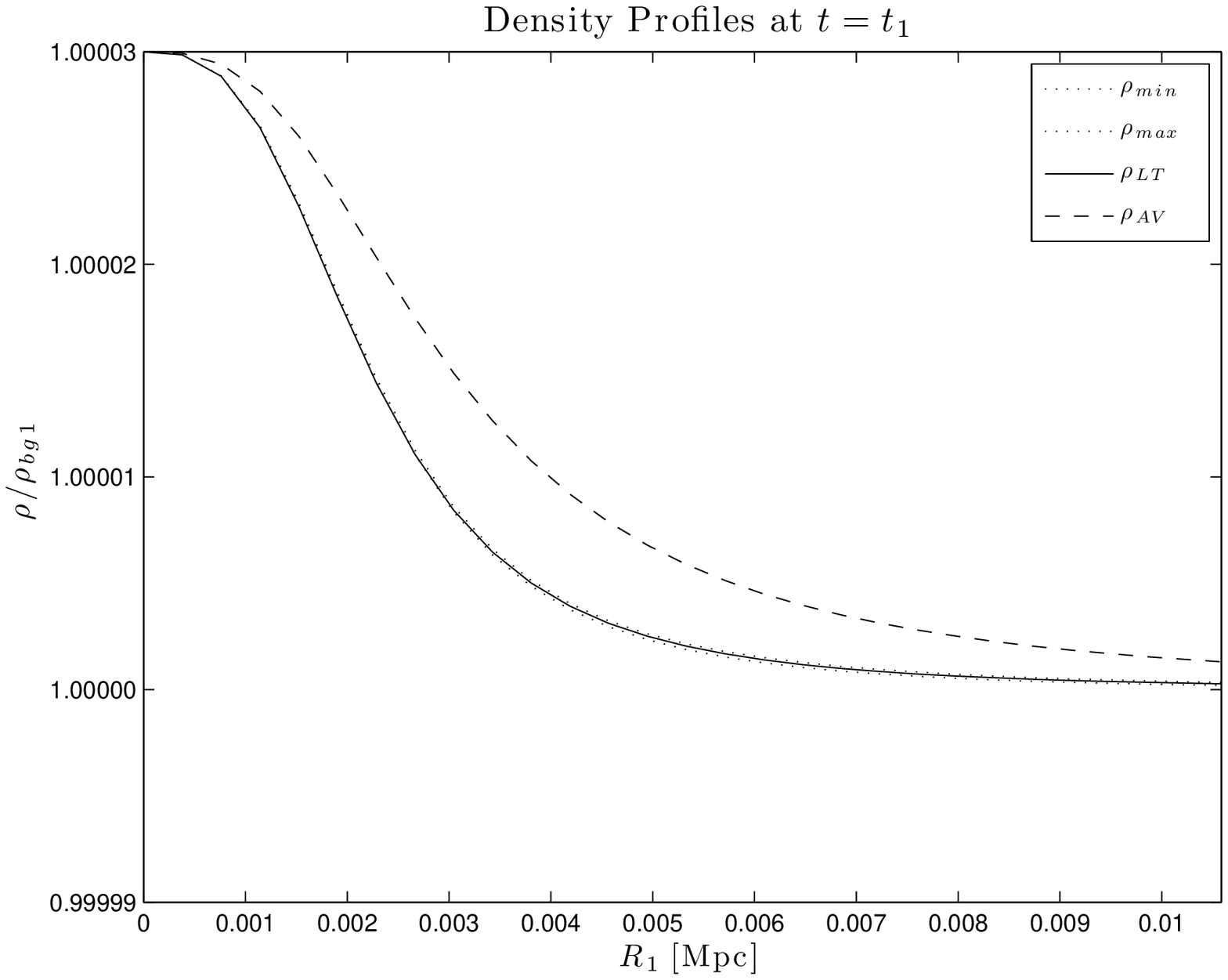}
\end{center}
\caption[Initial Density Profiles from Run \#2]{\textbf{Initial Density Profiles from Run \#2} - The minimum, maximum, internal average and LT density profiles at $t=t_1$ from Run \#2. Only LT-density was specified. The rest were calculated. (The profiles extend out to $R_1=0.27$) }
\label{run2_rho1}
\end{figure}

\begin{figure}
\begin{center}
\includegraphics[scale=0.6]{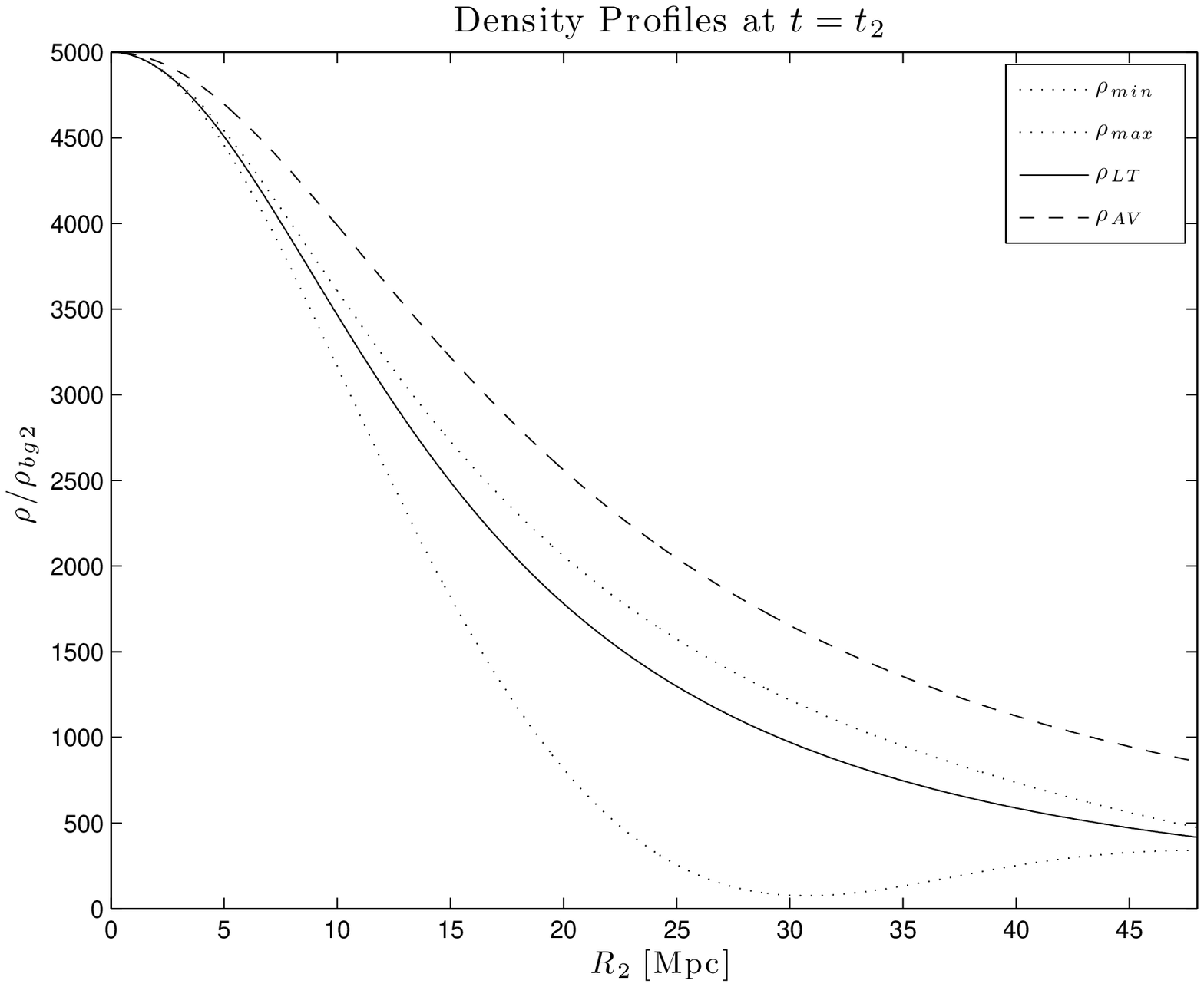}
\end{center}
\caption[Final Density Profiles from Run \#2]{\textbf{Final Density Profiles from Run \#2} - The minimum, maximum, internal average and LT density profiles at $t=t_2$ from Run \#2. The minimum and LT-density was specified. The rest were calculated.}
\label{run2_rho2}
\end{figure}

The density profiles $\rho_{LT}$, $\rho_{min}$, $\rho_{max}$ and $\rho_{AV}$ at the initial and final times are shown in Figures \ref{run2_rho1} and \ref{run2_rho2}, respectively. As in the case of Run \#1, both have an over-density at the origin, resulting in $\rho_{AV} > \rho_{LT}$ for all $(t,r)$, and thus the relationship between $E'/E|_{max}$ and $E'/E|_{\rho_{min}}$ is constant. In this case the $(\theta,\phi)$ orientation of the density dipole varies as R increases, describing one circuit of a helix over the range of the model. Again, the large deviation of $\rho_{min}$ away from $\rho_{LT}$, at the final time, causes little deviation of $\rho_{max}$ away from $\rho_{LT}$.  The function $E'/E|_{max}$, shown by the broken line in Figure \ref{run1_ere}, approaches a finite origin value, as expected from our choice of $\mu$. The broken lines in Figure \ref{run1_mftb} shows the LT arbitrary functions, $M$, $f$ and $t_b$, which are similar to those from Run \# 1. The Szekeres arbitrary functions, $S$, $P$ and $Q$, are shown by the broken lines in Figure \ref{run1_spq}. The effect of varying the dipole orientation angles is evident in all of them. Surfaces showing the full time evolution of $\rho_{LT}$, $\rho_{min}$ and $\rho_{max}$ are qualitatively very similar to those from Run \#1 (see Figures \ref{run1_rholt}, \ref{run1_rhomin} and \ref{run1_rhomax}), and are thus omitted here.

\subsubsection{Run \#3}
\begin{figure}
\begin{center}
\includegraphics[scale=0.6]{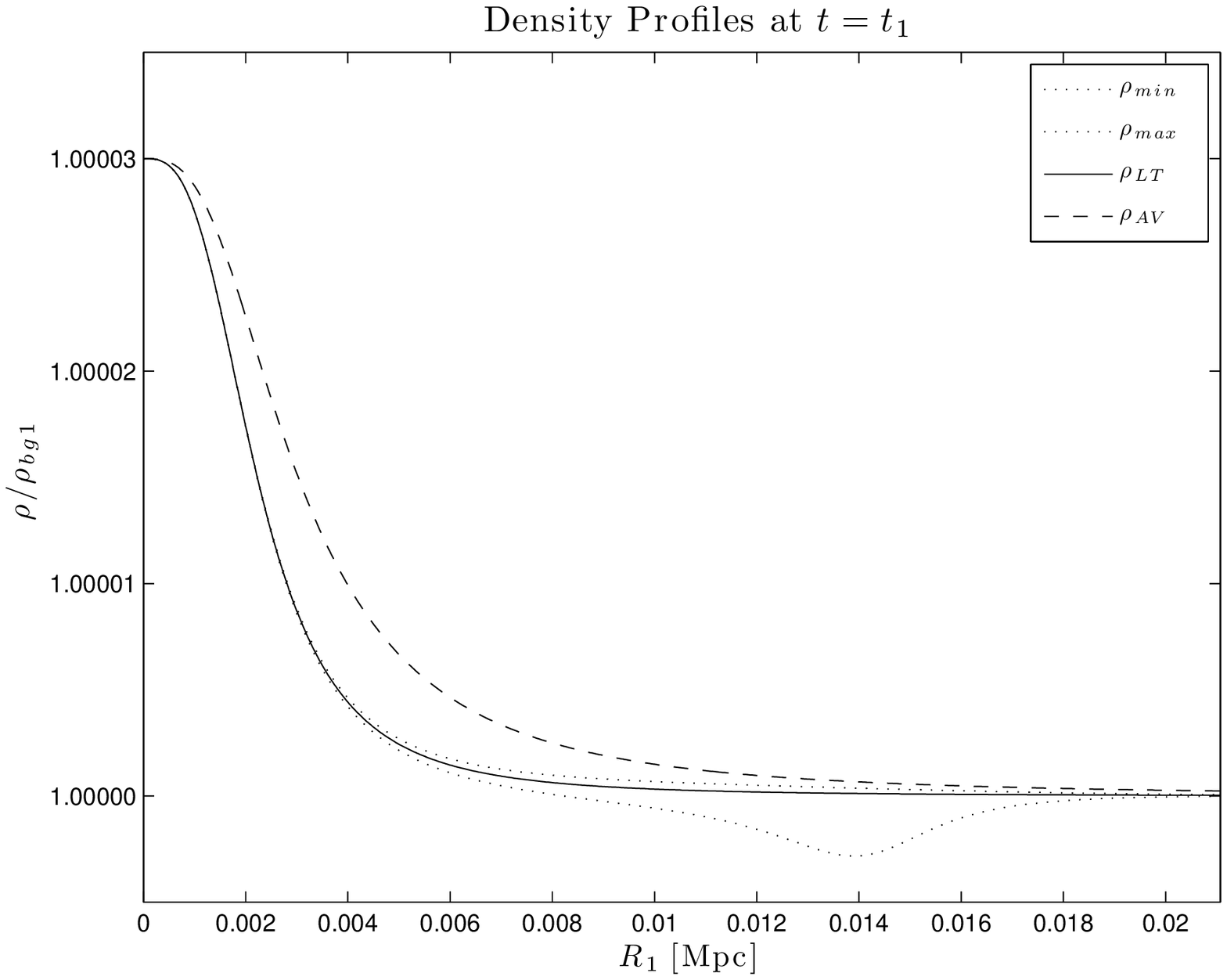}
\end{center}
\caption[Initial Density Profiles from Run \#3]{\textbf{Initial Density Profiles from Run \#3} - The minimum, maximum, internal average and LT density profiles at $t=t_1$  from Run \#3. Only LT-density was specified. The rest were calculated.}
\label{run3_rho1}
\end{figure}

\begin{figure}
\begin{center}
\includegraphics[scale=0.62]{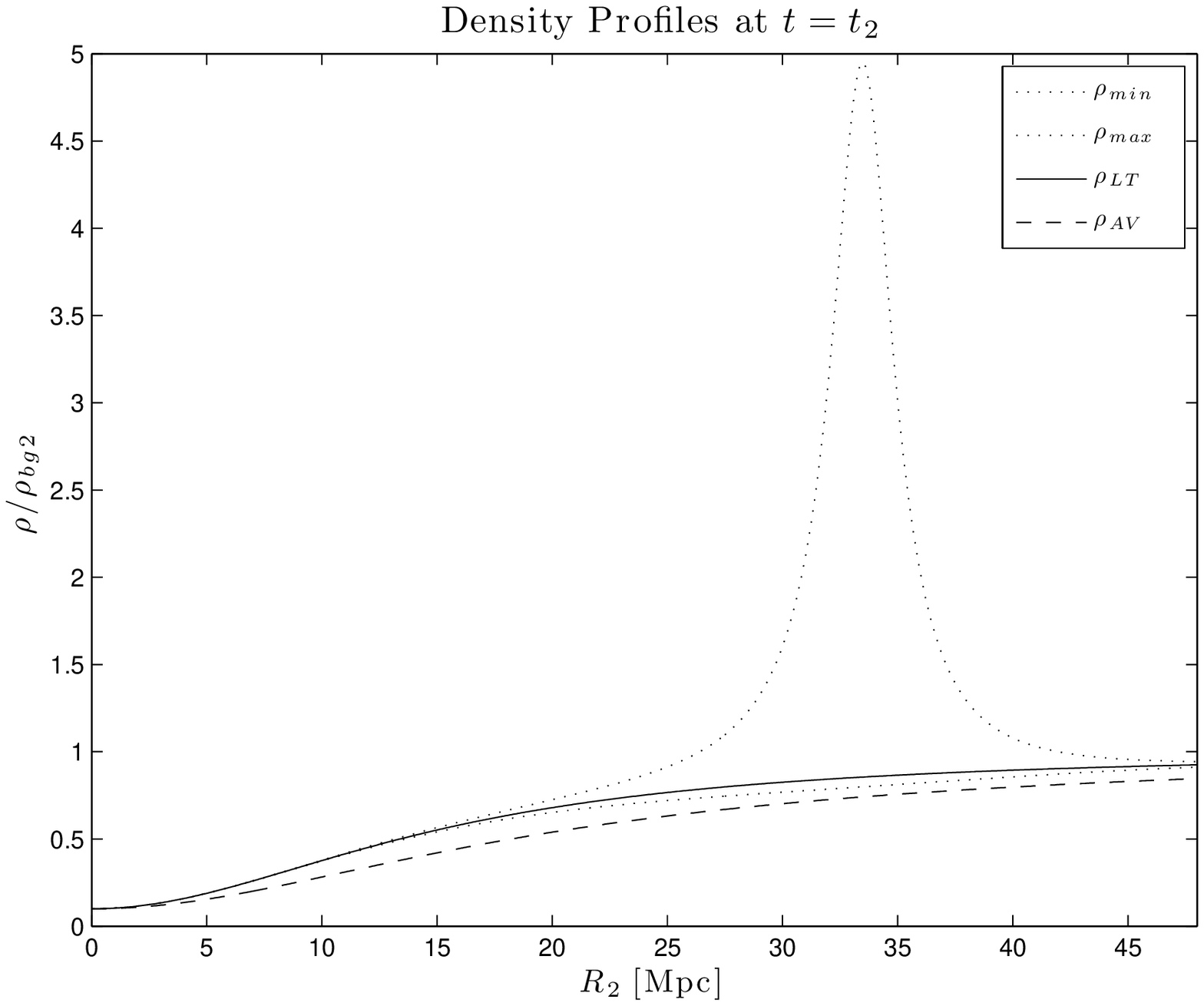}
\end{center}
\caption[Final Density Profiles from Run \#3]{\textbf{Final Density Profiles from Run \#3} - The minimum, maximum, internal average and LT density profiles at  $t=t_2$ from Run \#3. The minimum and the LT-density was specified at $t_1$ and $t_2$. The rest were calculated.}
\label{run3_rho2}
\end{figure}

\begin{figure}
\begin{center}
\includegraphics[scale=0.62]{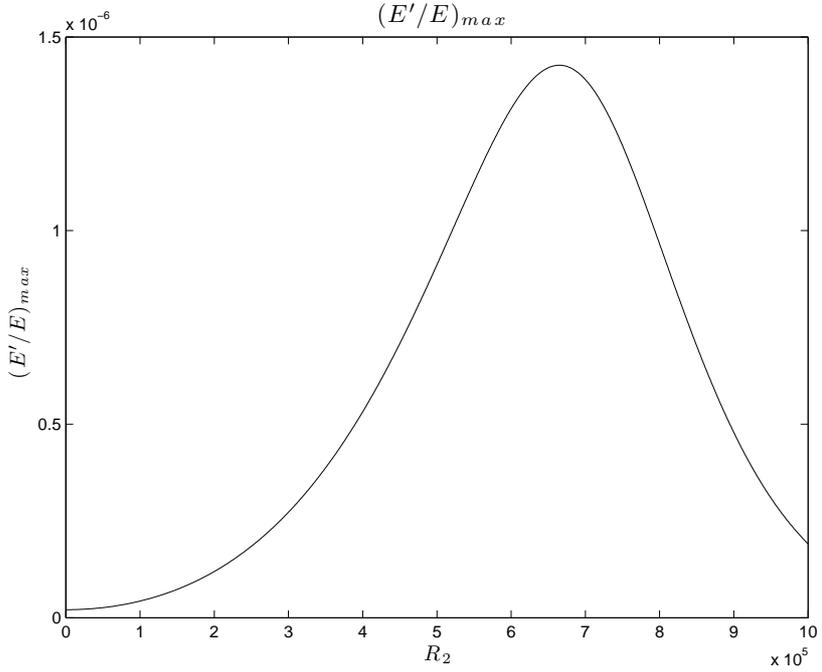}
\end{center}
\caption[The Function $E'/E|_{max}$ from Run \#3]{\textbf{The Function $E'/E|_{max}$ from Run \#3} - The dipole function $E'/E|_{max}$ from Run \#3}
\label{run3_ere}
\end{figure}

\begin{figure}
\begin{center}
\includegraphics[scale=0.8]{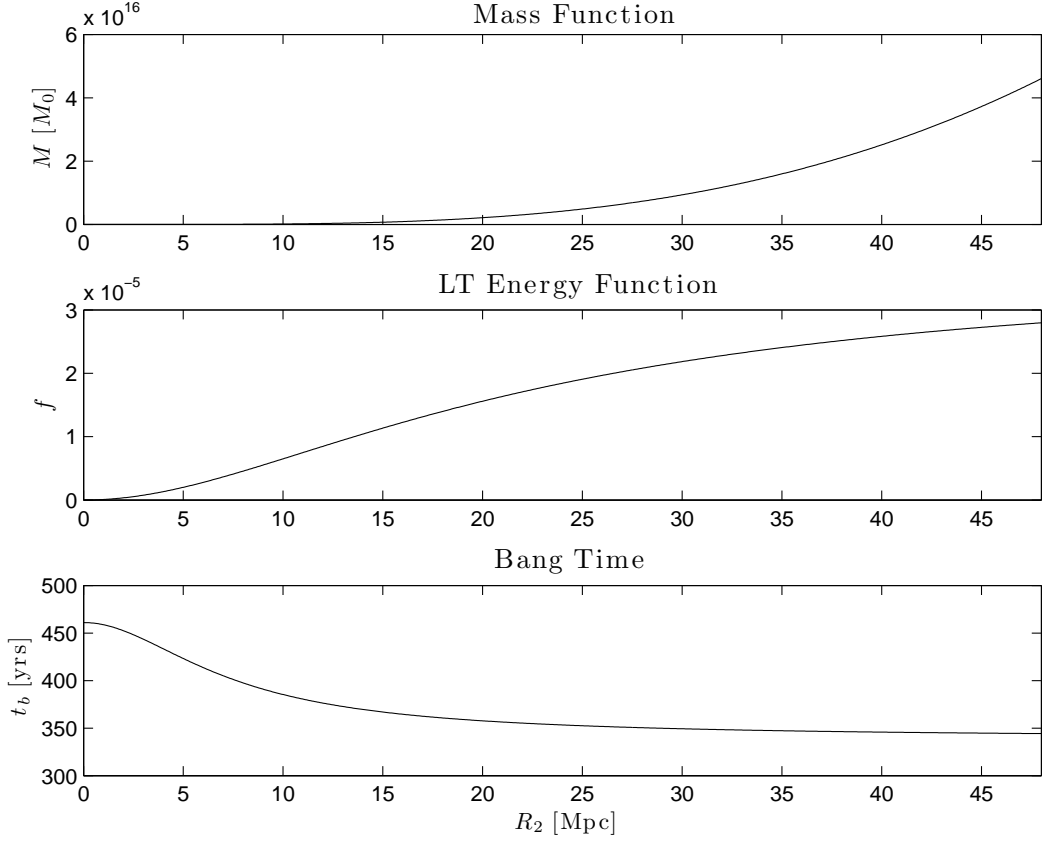}
\end{center}
\caption[Arbitrary Functions $M$, $f$ and $t_b$ from Run \#3]{\textbf{Arbitrary Functions $M$, $f$ and $t_b$ from Run \#3} - The arbitrary functions which are common to both Szekeres and LT modes; $M$, $f$ and $t_b$}
\label{run3_mftb}
\end{figure}

\begin{figure}
\begin{center}
\includegraphics[scale=0.8]{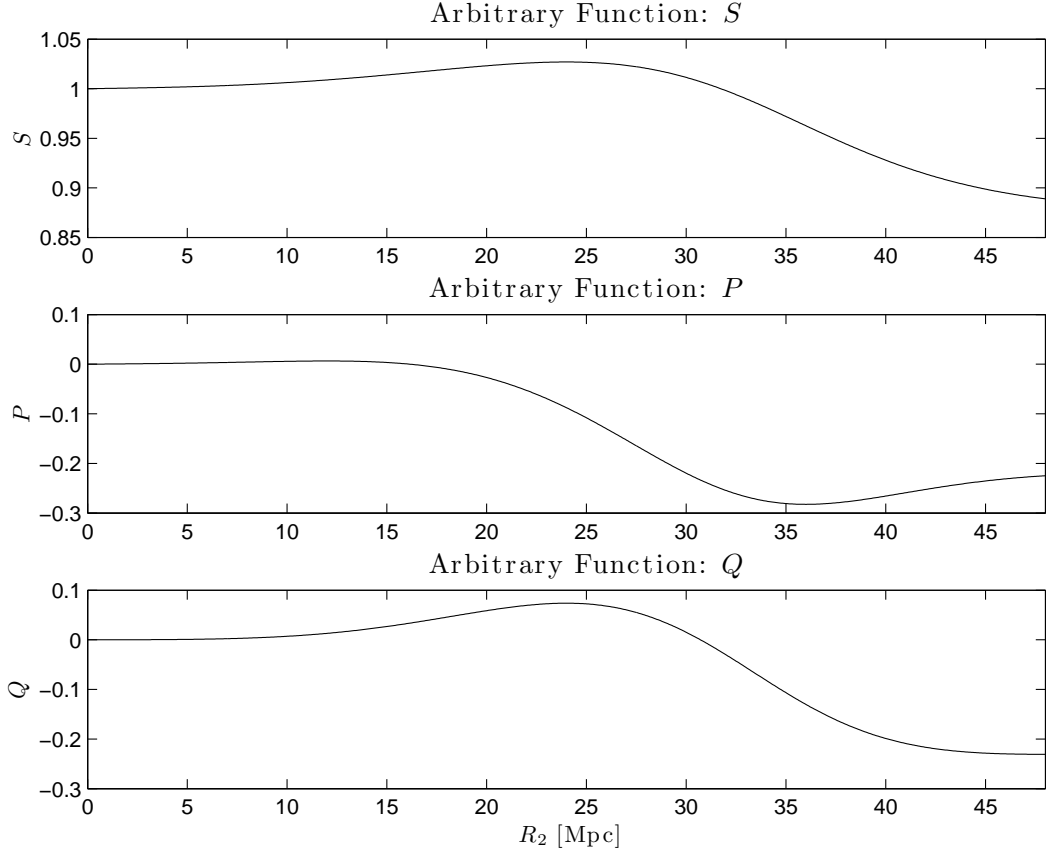}
\end{center}
\caption[Arbitrary Functions $S$, $P$ and $Q$ from Run \#3]{\textbf{Arbitrary Functions $S$, $P$ and $Q$ from Run \#3} - The arbitrary functions which are unique to Szekeres models; $S$, $P$ and $Q$}
\label{run3_spq}
\end{figure}

\begin{figure}
\begin{center}
\includegraphics[scale=0.6]{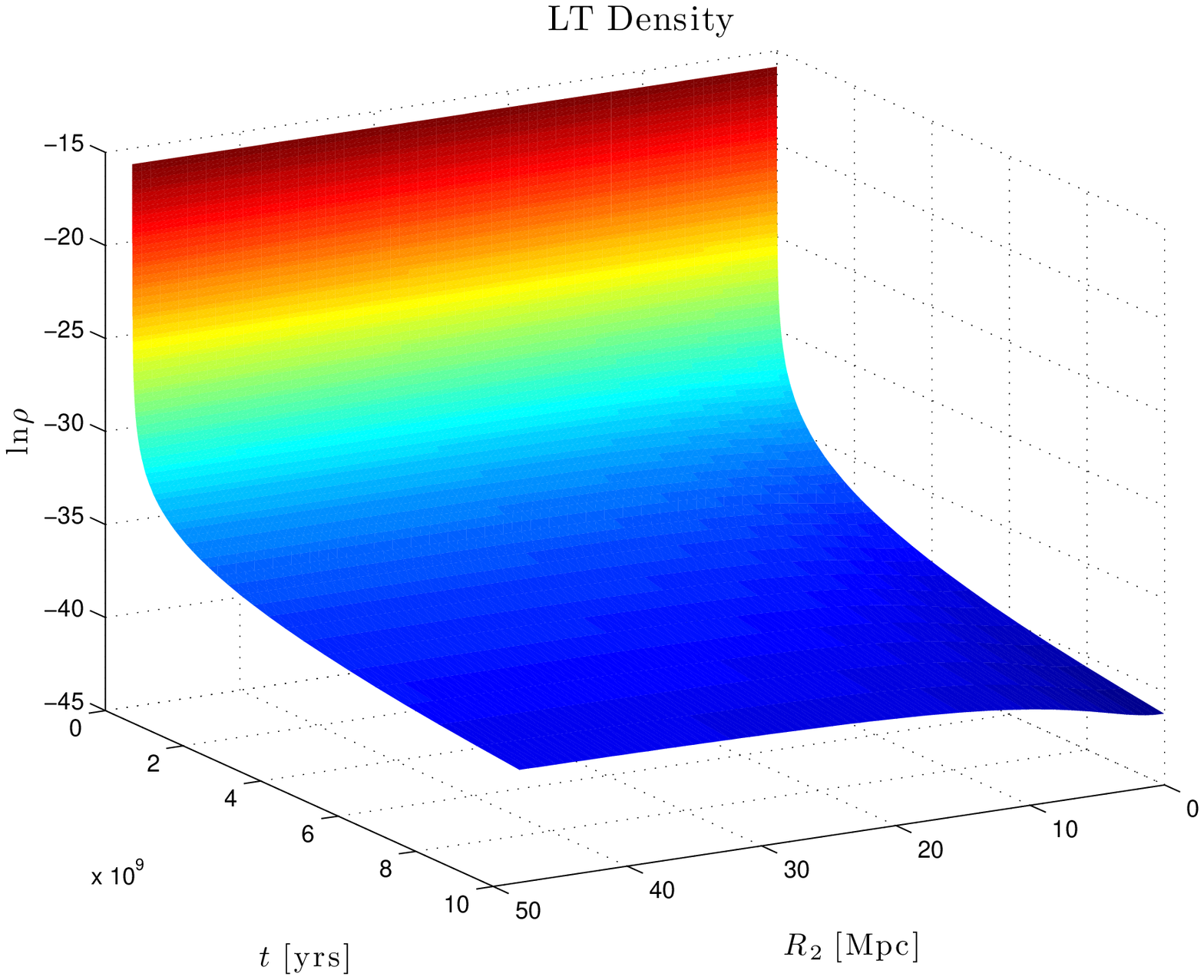}
\end{center}
\caption[Evolution of the LT-density from Run \#3]{\textbf{Evolution of the LT-density from Run \#3} - The evolution of the LT-density profile, $\rho_{LT}$. Log scale}
\label{run3_rholt}
\end{figure}

\begin{figure}
\begin{center}
\includegraphics[scale=0.6]{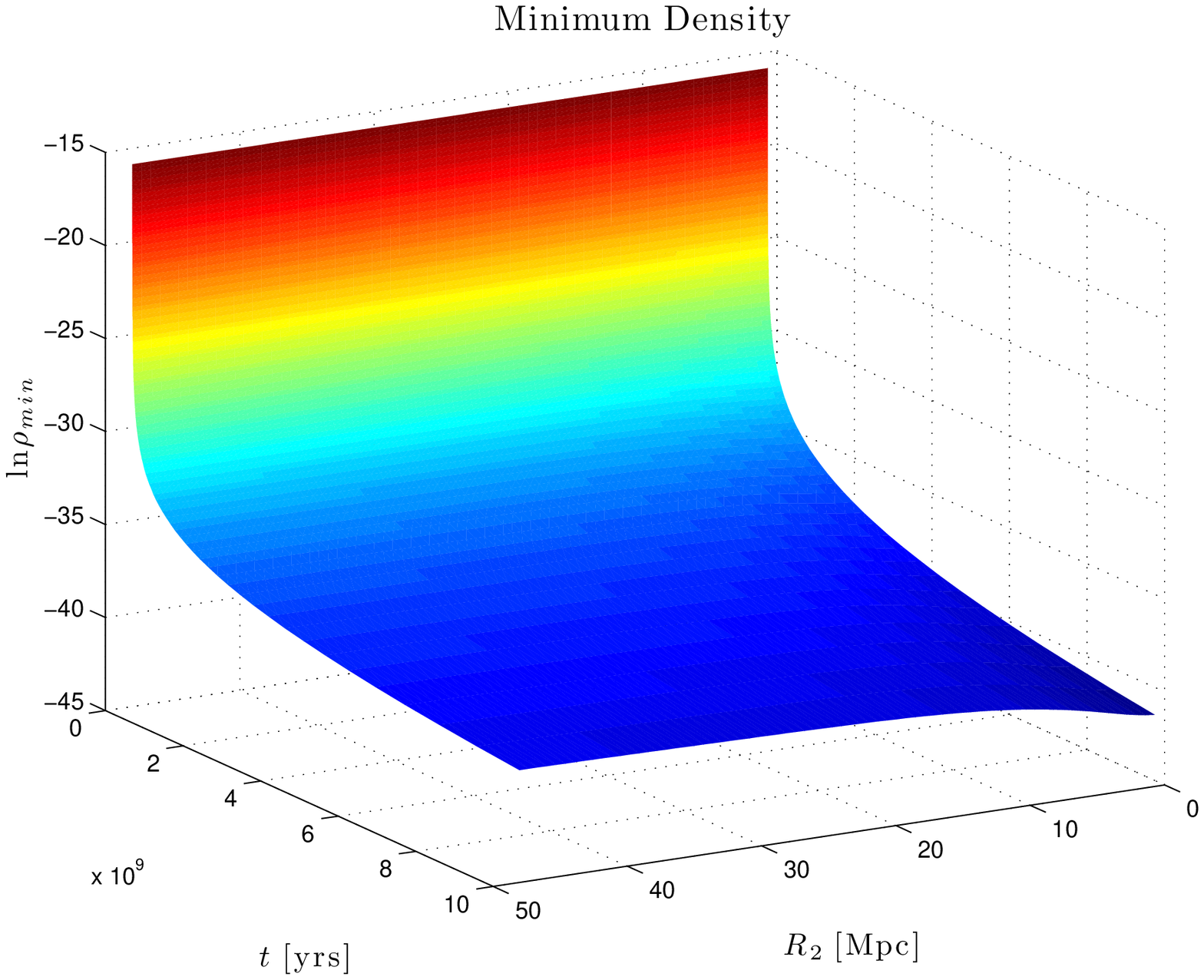}
\end{center}
\caption[Density Dipole Minimum from Run \#3]{\textbf{Density Dipole Minimum from Run \#3} - The evolution of the density dipole minimum, $\rho_{min}$. Log scale}
\label{run3_rhomin}
\end{figure}

\begin{figure}
\begin{center}
\includegraphics[scale=0.6]{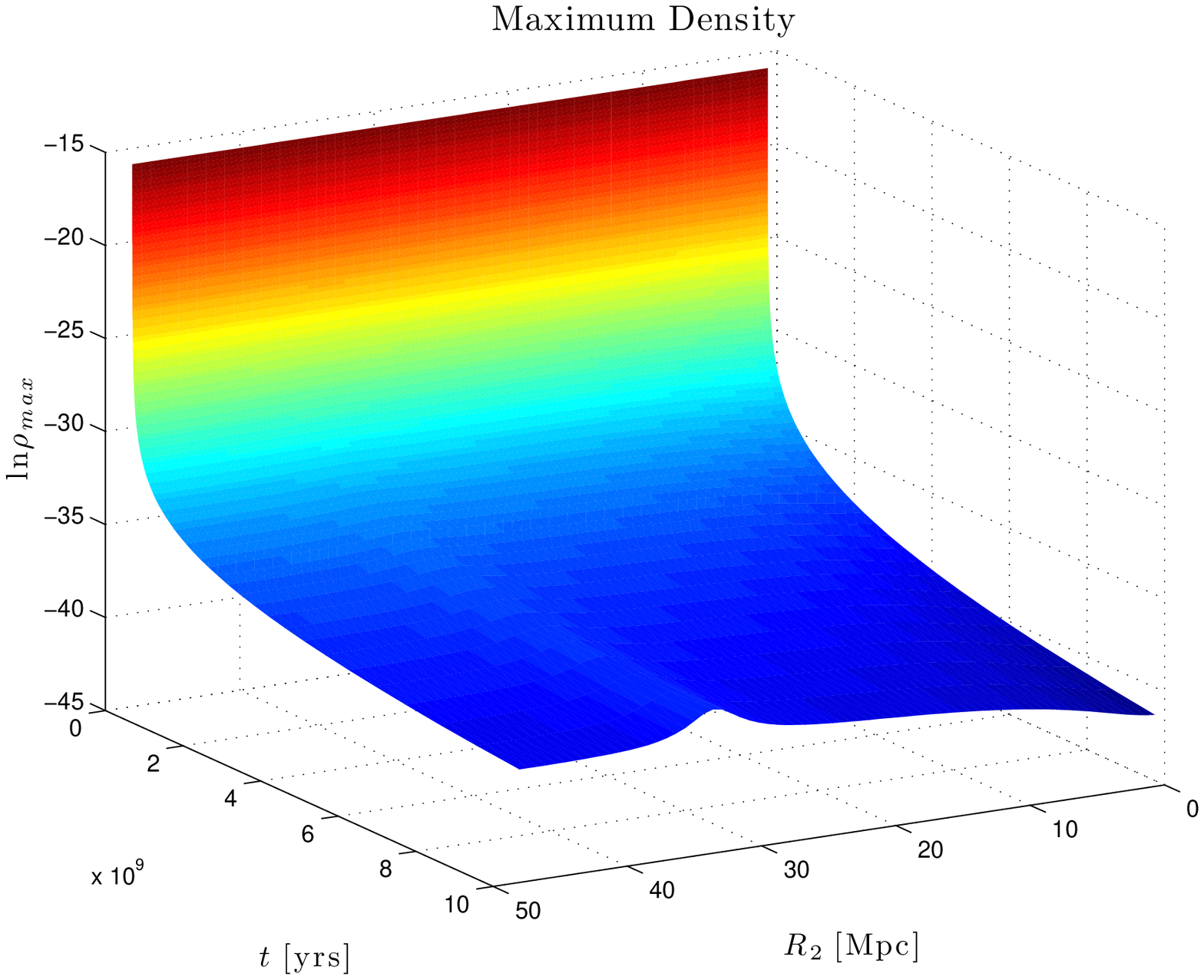}
\end{center}
\caption[Density Dipole Maximum from Run \#3]{\textbf{Density Dipole Maximum from Run \#3} - The evolution of the density dipole maximum, $\rho_{max}$. Log scale}
\label{run3_rhomax}
\end{figure}

The density profiles $\rho_{LT}$, $\rho_{min}$, $\rho_{max}$ and $\rho_{AV}$ at the initial and final times are shown in Figures \ref{run3_rho1} and \ref{run3_rho2}, respectively. Unlike the previous two cases, the initial over-density evolves into an under-density at the final time, causing the quantity $(\rho_{LT} - \rho_{AV})$ to change sign during the course of the evolution. Thus, the relationship between $E'/E|_{max}$ and $E'/E|_{\rho_{min}}$ is not constant for all $(t,r)$ - a `flip' takes place, occurring on a non-linear locus at or near $t_1$, and covering the whole range of $r$ within 1700 years. Though the density dipole orientation is as for Run \#2, the density dipole orientation is reversed at the initial time. At the final time, the `deviation function ' produces a large difference between $\rho_{LT}$ and $\rho_{max}$, but little between $\rho_{LT}$ and $\rho_{min}$. This effect seems to be depend on the sign of $(\rho_{LT} - \rho_{AV})$. If $R'(\rho_{LT} - \rho_{AV})>0$, then the deviation in maximum density is enhanced as compared to the minimum density, and vice versa. Furthermore, the location of the peak in $\rho_{max}$ corresponds to the location of the minimum in $\rho_{min}$, which, by definition, corresponds to the peak in the deviation function, $\mu$. The function $E'/E|_{max}$, shown in Figure \ref{run3_ere}, approached a finite origin value, as expected from our choice of $\mu$. Figure \ref{run3_mftb} shows the LT arbitrary functions, $M$, $f$ and $t_b$. The Szekeres arbitrary functions, $S$, $P$ and $Q$, are shown in Figure \ref{run3_spq}. Surfaces showing the full time evolution of $\rho_{LT}$, $\rho_{min}$ and $\rho_{max}$ are shown in Figures \ref{run3_rholt} \ref{run3_rhomin} and \ref{run3_rhomax}, respectively.

\section{Conclusions} 
\label{conclusions}

Modern cosmology, although having enjoyed a great number of successes in the last century, is still faced with many open questions (e.g. the apparent dimming of distant supernovae). This has given rise to many highly speculative theories receiving much of the spotlight in recent years. While the importance of such research cannot be discounted, we should not loose sight of the full implications of the best current theory, GR. The consequences of the non-linearity of the EFE's have not yet been fully explored in cosmology, and investigating exact solutions is an indispensable tool for doing this. In this respect, the Szekeres family of inhomogeneous solutions offers a wide range of possibilities for modelling cosmic structure. When attempting to construct a realistic model of the universe, or part thereof, it is of great utility to do so from physical quantities or data more directly accessible to observation than theoretical metric functions. In other words, there's a need to be able to create models that incorporate observational constraints from two different times, such as recombination and recent times.

We considered quasispherical Szekeres models, outlining a model construction procedure using given density data at some initial and final time. Szekeres models are thought to be a sequence of non-concentric mass shells, each with density dipole. Consequently, the procedure requires one to specify `radial' profiles of the equatorial density at the initial and final time, as well as dipole parameters (encoding the intensity and orientation), which we choose to specify at the final time. Since, in Szekeres models, the dynamics of the areal radius is identical to that in LT models, the evolution of each shell can be determined by the equatorial density profiles, $\rho_{LT}(t_1, R)$ and $\rho_{LT}(t_2, R)$, as is the case in LT models. We used a minor modification of the LT model construction procedure of Krasinski \& Hellaby to determine the arbitrary functions which are common to both - $M$, $f$ and $t_b$.  The dipole variations over each shell are encoded in the `deviation function', $\mu( R)$, that defines the minimum density $\rho_{min}$ on each $R$ shell, and the orientation angles, $\theta_{\rho_{min}}(R)$ and $\phi_{\rho_{min}}(R)$, of that minimum density. With knowledge of $\rho_{LT}(R)$ and $\mu(R)$ one can determine $\rho_{min}(R)$ and $\rho_{max}(R)$. For the determination of the metric functions unique to Szekeres models, $S$, $P$ and $Q$, representing degrees of freedom pertaining to the dipole, we derive a new result - exact analytic expressions in terms of the dipole parameters, $\theta_{\rho_{min}}$, $\phi_{\rho_{min}}$ and $E'/E|_{max}$ (which is directly related to $\rho_{min}$ and $\rho_{LT}$). So, by specifying the profiles $\rho_{LT}(t_1, R)$, $\rho_{LT}(t_2, R)$, $\mu(R)$, $\theta_{\rho_{min}}(R)$ and $\phi_{\rho_{min}}(R)$, and following the algorithm given in \S4.5, one can determine the six arbitrary metric functions which completely define the Szekeres model. Special attention was paid to the origin limit of the dipole parameter $E'/E$, investigating the conditions which ensure it is finite and non-zero there. We corrected a claim made in previous literature that a maximum in $E'/E$ corresponds to a density minimum (see Equation 69 in \cite{Hellaby:2002nx}) by showing that the derivative of the density with respect to $E'/E$ is not always negative, but rather, it depends on the sign of $R'( \rho_{LT} -\rho_{AV})$. We found that a maximum in $E'/E$ corresponds to a density minimum if $R'( \rho_{LT} -\rho_{AV})<0$.

Using MATLAB, code was written to implement the procedure for determining these metric functions, as well as to simulate the model evolution. Since investigating elaborate models is beyond the scope of this work, we considered only three simple cases. All models spanned the time from recombination until the present, with the choice of initial density profile consistent with CMB anisotropies. We then chose different profiles and orientation angles at the final time for each of the models. In all cases, when reconstructing the model evolution, the calculated metric functions reproduced the specified initial and final density profiles. In Runs \#1 and \#2  the initial and final profiles both have an over-density at the origin, causing the sign of $(\rho_{LT}-\rho_{AV})$ to be negative for all $(t,r)$, and hence $E'/E|_{max}$ occurs at density minimum for all $(t,r)$. In both cases, the deviation of $\rho_{min}$ away from $\rho_{LT}$ was much greater than that in $\rho_{max}$.  In contrast, Run \#3 had an initial over-density evolving into an under-density at the final time, which caused $(\rho_{LT}-\rho_{AV})$ to change sign during the course of the model evolution. Also, the deviation in $\rho_{min}$ away from $\rho_{LT}$, at the final time, is much less than the deviation in $\rho_{max}$. We speculate that the profile which shows the biggest deviation away from $\rho_{LT}$, either  $\rho_{min}$ or $\rho_{max}$, seems to depend on the sign of $(\rho_{LT}-\rho_{AV})$.

In the future, plotting the density on a section of geodesic spatial 2-surfaces will give one a better feel for the resulting density distribution. Since the constant-$(t,r)$ mass shells are interpreted as being arranged non-concentrically, the distance from the origin to a given shell is not trivially related to areal radius, but instead, has $(\theta,\phi)$ variations. Hence, plotting $\rho_{min}$, $\rho_{max}$ and $\rho_{LT}$ as a function of $r$, whatever one's choice for $r$ may be, does not give one a sense of the density profile in any direction. In addition, the local $(\theta,\phi)$ coordinates, which describe the dipole orientation on each shell, are not necessarily parallel. In order to understand the relationship between the coordinates on each shell some parallel transport operation may need to be performed. It is also foreseeable that, in the future, the model construction algorithm could be extended to include velocity profiles, as was done in \cite{Krasinski:2003yp} for the LT model. Extending these results to the $\epsilon \neq 1$ Szekeres cases will require care as the interpretation of $R$ and $M$ is different from the quasi-spherical case \cite{Hellaby:2007hq}.

\appendix
\section{Details of the LT model construction}
\label{appendix}
Here we provide some important equations from the LT model construction procedure of \cite{Krasinski:2001yi}, in which the evolution is determined from initial and final density profiles, though adapted to the coordinate choice (\ref{r=R}). In particular, we list the cases that arise, their ranges of validity, and the equations to be solved in each case. This is a core part of the proposed Szekeres model construction procedure presented in \S\ref{work}.

\subsection{Finding \boldmath $f$ and $t_b$}
This material follows on from (\ref{M(R)}) of section \ref{LTmodelconstr}.

With the variables\footnote{$a$ and $x$ have the advantage of being non-zero at the origin}
\begin{align}
a_i = \frac{R_i}{M^{1/3}}, \hspace{1cm} x = \frac{|f|}{M^{2/3}},	\label{ax}
\end{align}
the nature of the LT model that evolves between the initial and final time slice at a given $M$ is\\
{\bf Hyperbolic} ($f>0$)\\
if
\begin{equation}
 t_2 - t_1 < \frac{\sqrt{2}}{3}(a_2^{3/2} - a_1^{3/2}), \label{ineq1}
\end{equation}
then, the energy function is given by
\begin{equation}
 f = xM^{2/3}, \label{f1}
\end{equation}
and the bang time by 
\begin{align}
t_b = t_i - x^{-3/2}\left[ \sqrt{(1+xa_i)^2-1} - \operatorname{arcosh}(1+a_ix)\right] \label{tb1}
\end{align}
where $x$ solves
\begin{align}
 0=\psi_{HX}(x) =&\sqrt{(1+a_2x)^2-1}-\operatorname{arcosh}(1+a_2x)\notag\\
 &-\sqrt{(1+a_1x)^2-1}+\operatorname{arcosh}(1+a_1x)-(t_2-t_1)x^{3/2}. \label{psix1}
\end{align}

\bigskip\noindent {\bf Near Parabolic} ($f \approx 0$)\\
 if ($t_2-t_1$) is close to
\begin{equation}
 t_2 - t_1 = \frac{\sqrt{2}}{3}(a_2^{3/2} - a_1^{3/2}), \label{ineq2}
\end{equation}
then, the energy function is given by
\begin{equation}
 f = xM^{2/3}, \label{f2}
\end{equation}
and the bang time by
\begin{equation}
 t_b = t_i - a^{3/2}_i\left(1 - \frac{3}{20}a_ix + \frac{9}{224}a_i^2x^2 \right), \label{tb2}
\end{equation}
where $x$ solves
\begin{align}
 0=\psi_P(x) \approx \frac{\sqrt{2}x^{3/2}}{3}\bigg[&a^{3/2}_2\left(1 - \frac{3}{20}a_2x + \frac{9}{224}a_2^2x^2 \right) \notag\\
 &- a^{3/2}_1\left(1 - \frac{3}{20}a_1x + \frac{9}{224}a_1^2x^2 \right) - (t_2 - t_1) \bigg]. \label{psix2}
\end{align}

\bigskip\noindent {\bf Elliptic and still expanding at $t_2$} ($f<0$)\\
if
\begin{equation}
(a_2/2)^{3/2} \left[\pi - \arccos(1-2a_1/a_2) +2\sqrt{a_1/a_2 - (a_1/a_2)^2} \right]  \geq t_2 - t_1 > \frac{\sqrt{2}}{3}(a_2^{3/2} - a_1^{3/2}), \label{ineq3}
\end{equation}
then, the energy function is given by
\begin{equation}
 f = -xM^{2/3}, \label{f3}
\end{equation}
and the bang time by 
\begin{equation}
 t_b = t_i - x^{-3/2}\left[\arccos(1-a_ix)-\sqrt{1-(1-a_ix)^2}\right], \label{tb3}
\end{equation}
where $x$ solves
\begin{align}
  0=\psi_X(x) =&\arccos(1-a_2x)- \sqrt{1 - (1-a_2x)^2} \notag\\
  &-\arccos(1-a_1x) +\sqrt{1-(1-a_1x)^2} -(t_2-t_1)x^{3/2}. \label{psix3}
\end{align}

\bigskip\noindent {\bf Elliptic and near maximum expansion at $t_2$} ($f<0$)\\
if ($t_2-t_1$) is close to
\begin{equation}
t_2 - t_1 = (a_2/2)^{3/2} \left[\pi - \arccos(1-2a_1/a_2) +2\sqrt{a_1/a_2 - (a_1/a_2)^2} \right], \label{ineq4}
\end{equation}
then, then energy function is given by 
\begin{equation}
 f = -xM^{2/3}, \label{f4}
\end{equation}
and the bang time by
\begin{align}
 t_b &= t_1 - x^{-3/2}\left[\arccos(1-a_1x)-\sqrt{1-(1-a_1x)^2}\right]\notag\\
&\approx t_2 - x^{-3/2}\left[ \pi - 2^{3/2}(2-a_2x)^{1/2} + \frac{2^{3/2}}{12}(2-a_2x)^{3/2}  \right], \label{tb4}
\end{align}
where $x$ solves
\begin{align}
  0=\psi_M(x) =&\pi -2^{3/2}(2-a_2x)^{1/2} + \frac{2^{3/2}}{12}(2-a_2x)^{3/2}  \notag\\
  &-\arccos(1-a_1x) +\sqrt{a_1x(2-a_1x)} -(t_2-t_1)x^{3/2}. \label{psix4}
\end{align}

\bigskip\noindent {\bf Elliptic and recollapsing at $t_2$} ($f<0$)\\
if
\begin{equation}
t_2 - t_1 > (a_2/2)^{3/2} \left[\pi - \arccos(1-2a_1/a_2) +2\sqrt{a_1/a_2 - (a_1/a_2)^2} \right], \label{ineq5}
\end{equation}
then, the energy function is given by
\begin{equation}
 f = -xM^{2/3}, \label{f5}
\end{equation}
and the bang time by 
\begin{align}
 t_b &= t_1 - x^{-3/2}\left[\arccos(1-a_1x)-\sqrt{1-(1-a_1x)^2}\right]\notag\\
&= t_2 - x^{-3/2}\left[ \pi + \arccos(-1+a_2x) + \sqrt{1 - (1-a_2x)^2}  \right], \label{tb5}
\end{align}
where $x$ solves
\begin{align}
  0=\psi_C(x) =& \pi + \arccos(-1+a_2x) + \sqrt{1 - (1-a_2x)^2} \notag\\
  &-\arccos(1-a_1x) +\sqrt{1-(1-a_1x)^2} -(t_2-t_1)x^{3/2}. \label{psix5}
\end{align}

\subsection{Solving \boldmath $\psi(x)=0$ }
The value of $x$ which solves $\psi(x)=0$, at each $M$ value,  can be found numerically using the bisection method. The range in $x$ over which to bisect and a good starting value for the first guess, $x_g$, for each evolution type, are give in Table \ref{bisect}.\\

\begin{table}[h]
\centering
\begin{tabular}{|r|c|c|}
\hline Evolution Type & Bisection Range \\ 
\hline \hline HX & $0..( \frac{a_2-a_1}{t_2-t_1} )^2$\\ 
\hline all other types & $0..\frac{2}{a_2}$ \\ 
\hline 
\end{tabular} 
\caption[Bisection Info]{\textbf{Bisection Info} - Showing, for various evolution types, the range for the bisection method which is used to solve the various $\psi(x)=0$ equations.}
\label{bisect}
\end{table}

Increasing numerical error in the exact expressions for $\psi(x)$ necessitates the use of series expansions, as the borderline cases are approached. Thus, `fat' borderline equations are used when solving for $x$ in regions near the PX or EM cases. The range over which these `fat' borderline expressions are valid is assumed to be, approximately, the region where the ratio of the magnitude of the third order component of the series expansion, to that of the first, is less than $10^{-3}$.

\subsection{Limiting values at \boldmath{$M=0$}}
While several quantities have the value $0$ at the origin, the variables $a_i$ and $x$ have finite limits as $M \rightarrow 0$. The origin value of $a$ is \cite{Krasinski:2001yi}
\begin{align}
a_i(0) =  \left( \frac{6}{\kappa \rho_i(0)} \right)^{1/3}, \label{ai0}
\end{align}
and the value of $x$, calculated by solving the relevant $\psi(x)=0$ equation, comes out non-zero automatically when non-zero values of $a_i(0)$ are used.

\subsection{Reconstructing model evolution}
When reconstructing the evolution of the model, it is convenient to re-write the LT solutions in terms of $a$ and $x$. The time evolution of the quantity $a$, for the various evolution types, is given below.\\

\noindent Elliptic:
\begin{align}
a= \frac{1-\cos\eta}{x}, \hspace{0.5cm} t - t_b = \frac{(\eta-\sin\eta)}{x^{3/2}}
 \label{RevolE}
\end{align}
\noindent Parabolic or close to it:
\begin{align}
a = \left(\frac{9}{2} \right)^{1/3} (t - t_b)^{2/3} \left( 1 + \frac{x}{20}[6(t-t_b)]^{2/3} - \frac{3x^2}{2800}[6(t-t_ b)]^{4/3} + \frac{23x^3}{504000}[6(t-t_b)]^2  \right)
\label{RevolP}
\end{align}
\noindent Hyperbolic:
\begin{align}
a = \frac{\cosh\eta-1}{x}, \hspace{0.5cm} t - t_b = \frac{(\sinh\eta-\eta)}{x^{3/2}}
 \label{RevolH}
\end{align}
In all cases the density is given by
\begin{align}
\rho = \frac{1}{4\pi a^2(a/3 + M a,_M)} \label{rholta}
\end{align}

\bibliographystyle{JHEP}
\bibliography{references}

\end{document}